%% file: main.tex
\documentclass[]{spie}  

\usepackage{amsmath,amsfonts,amssymb}
\usepackage{graphicx}
\usepackage[colorlinks=true, allcolors=blue]{hyperref}
\usepackage{tikz}
\usepackage{xcolor}

\usetikzlibrary{positioning,arrows.meta}

\title{Overview of the Canadian Hydrogen Observatory and Radio Transient Detector (CHORD) Project}
\input{chord_spie_authors_affiliations}

\authorinfo{Further author information: (Send correspondence to M.D.: E-mail: Matt.Dobb@McGill.ca)}

\begin{document} 
\maketitle

\newcommand{\ask}[2]{\textcolor{red}{\textbf{[Assigned to #1:} #2\textbf{]}}}
\newcommand{\missing}[2]{\textcolor{red}{\textbf{[Assigned to #1:} #2\textbf{]}}}
\newcommand{\placeholderfig}[1]{\fbox{\parbox{0.9\linewidth}{\centering #1}}}

\newcommand{\blue}[1]{\textcolor{blue}{(#1)}}

\begin{abstract}

The Canadian Hydrogen Observatory and Radio-transient Detector (CHORD)
is a next-generation wideband radio interferometer currently being
constructed and commissioned at the Dominion Radio Astrophysical Observatory in British
Columbia, Canada. CHORD is designed for precision 21\,cm cosmology,
fast radio transient discovery, spectral line galaxy surveys, and pulsar science using a highly
redundant large-N, small-diameter drift-scan array architecture. The
telescope consists of a 512-element core array of 6\,m dishes
operating from 300--1500\,MHz in drift-scan mode, together with two 64-dish outrigger
stations located at the Hat Creek Radio Observatory and the Green Bank
Observatory for long-baseline transient localization. The instrument
supports multiple simultaneous digital backends for interferometric
correlation, FRB detection, pulsar beamforming, and high spectral
resolution surveys. 
CHORD is designed with an emphasis on precision beam control and stable instrumental response, incorporating lessons learned from the Canadian Hydrogen Intensity Mapping Experiment (CHIME) while providing a substantial increase in sensitivity. 
Initial performance has been evaluated using a three-dish engineering array, and a 64-dish pathfinder array is currently being commissioned. The full array will be commissioned in 2028.

\end{abstract}




\section{Introduction and Project Overview}
\label{sec:intro}

The Canadian Hydrogen Observatory and Radio-transient Detector (CHORD) is a next-generation wideband radio interferometer based on a highly redundant large-N, small-diameter drift-scan array architecture. CHORD follows the remarkable scientific success of the Canadian Hydrogen Intensity Mapping Experiment (CHIME)~\cite{CHIMEOverview2022}, which established the large discovery space available for fast radio bursts (FRBs)\cite{CHIMEFRB2019} and has transformed the observational study of FRBs through the discovery of thousands of bursts and hundreds of repeating sources.\cite{CHIMECatalog1,CHIMECatalog2} CHIME has also recently achieved the first detection of the cosmological 21\,cm intensity mapping signal in auto-correlation at redshift $z\sim1$,\cite{CHIMEAuto21cm2025} demonstrating the scientific power of wide-field drift-scan radio interferometers for precision cosmology. CHORD builds directly on the scientific and technical lessons learned from CHIME, while extending the instrument concept with higher sensitivity, broader bandwidth, and improved control of instrumental systematics.

CHORD is designed simultaneously as a precision 21\,cm cosmology instrument, a deep fast-radio-transient discovery instrument, a pulsar survey telescope, and a wide-field radio survey instrument. 
A central motivation for the project is the recognition that many of the key challenges for 21\,cm cosmology are driven not by raw collecting area and sensitivity, but by precise control of instrumental response. 
Experience from CHIME demonstrated that beam precision and calibration redundancy/stability are critical for foreground subtraction and suppression of spectral systematics~\cite{CHORDWhitepaper2019}. 
These lessons have strongly shaped the CHORD design philosophy.


CHORD is the largest radio telescope ever constructed in Canada.
The CHORD core array consists of 512 six-meter dishes arranged in a highly redundant array geometry and located adjacent to CHIME at the Dominion Radio Astrophysical Observatory (DRAO) near Penticton, British Columbia. 
It operates in the 300--1500 \, MHz band using ultra-wideband feeds, low-noise integrated amplifiers, FPGA-based F-engines, GPU-based X-engines, and realtime software pipelines feeding several application-specific backends. 
Two additional outrigger stations, each with 64 six-meter dishes, are located at the Hat Creek Radio Observatory (HCO) and the Green Bank Observatory (GBO) adjacent to the CHIME outriggers~\cite{Amiri_2025}.
They provide continental baselines for transient localization and 
Very Long Baseline Interferometry (VLBI) applications. 
The telescope operates in drift-scan mode with no continuously moving mount systems, preserving the operational simplicity and survey speed demonstrated by CHIME.


Much of the CHORD instrument has been designed and prototyped directly within academic groups. 
This vertically integrated development model allows rapid iteration between science requirements, hardware implementation, firmware development, and commissioning strategy. 
The project combines contributions from universities, national laboratories, and industrial collaborators across Canada and internationally.

CHORD has been developed to optimize its science outcomes within a fixed funding envelope, rather than designing it against fixed, top-level requirements. 
This approach required continual tradeoffs between collecting area, bandwidth, correlator capacity, calibration requirements, infrastructure cost, and deployment complexity. 
Many major subsystems of the telescope were therefore optimized simultaneously for scientific performance, manufacturability, operational simplicity, and long-term maintainability. 
As the project was conceived and proposed pre-COVID pandemic, then implemented and built post-COVID, system optimization was ongoing to account for major changes in labor cost, raw materials and supply chain logistics, see Ref.\cite{DallasSPIE2026ProjManagement} for more info.


\begin{figure*}[t]
\centering
\includegraphics[width=\textwidth]{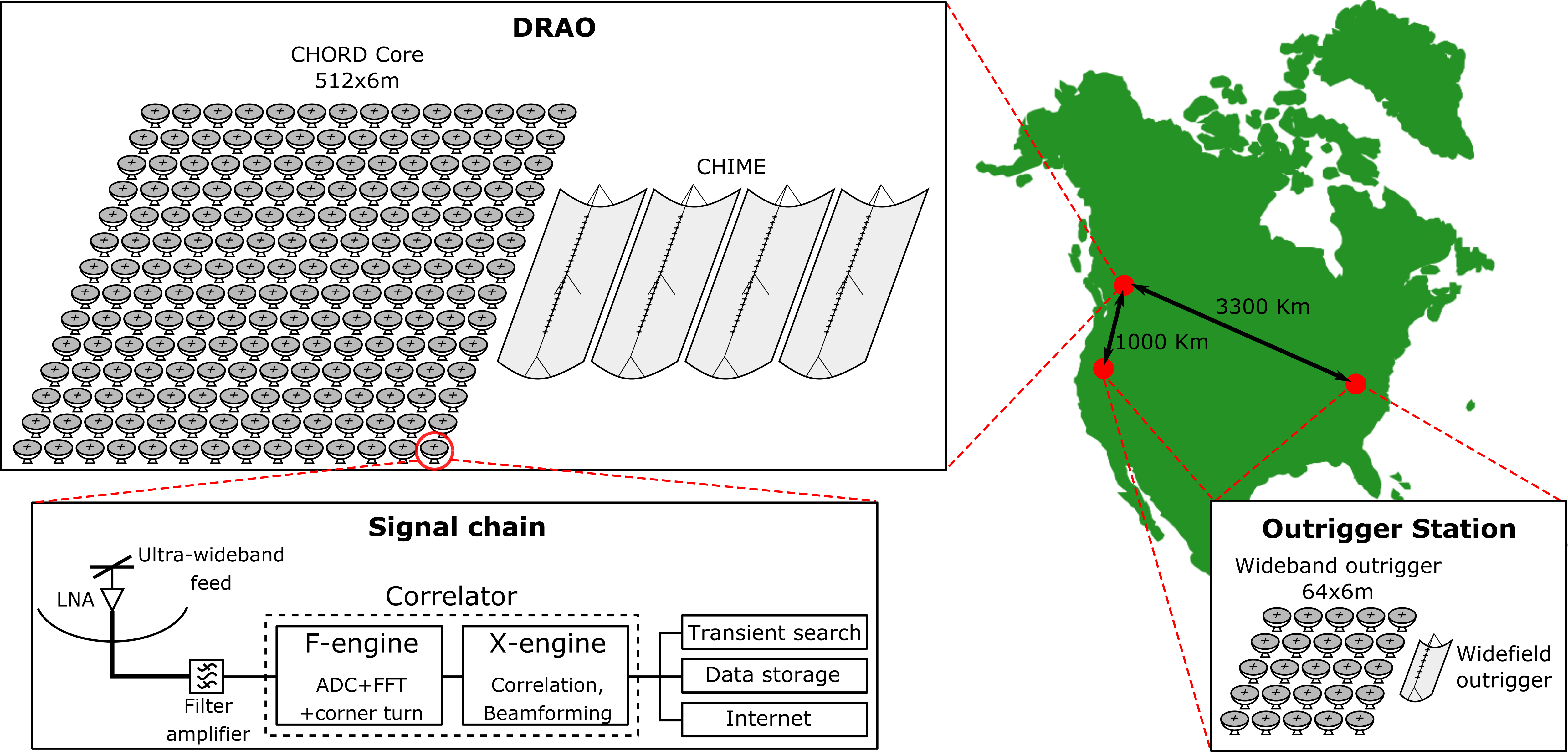}
\caption{
Overview of the CHORD observatory architecture. The central CHORD core
array located at the Dominion Radio Astrophysical Observatory (DRAO)
consists of 512 six-meter dishes operating over the 300--1500\,MHz
band. Two additional outrigger stations located at the Hat Creek Radio
Observatory (HCO) and the Green Bank Observatory (GBO) each contain
64 dishes and provide long baselines for precision transient localization
and VLBI applications. The lower panel illustrates the high-level signal
processing chain, including the analog receiver system, F-engine,
X-engine, and multiple commensal realtime processing backends supporting
FRB searches, pulsar beamforming, 
spectral line galaxy surveys, 
and 21\,cm cosmology observations.
}
\label{fig:chord_overview}
\end{figure*}









\section{CHORD Science Drivers}
\label{sec:science}






CHORD is designed as a commensal survey instrument.
The science goals
directly drive the telescope architecture, including the choice of
a highly redundant drift-scan array, ultra-wide instantaneous bandwidth, precision reflector surfaces, and multiple simultaneous digital backends.

\subsection{Fast Radio Transients}

CHORD is designed as a next-generation fast radio transient survey
instrument building directly on the discoveries of CHIME/FRB, which demonstrated both the ubiquity and diversity of the fast radio burst
population.
Relative to present-day CHIME, CHORD provides approximately twice the collecting
area, three times the instantaneous bandwidth, and substantially
lower system noise temperature. Combined, these improvements are expected
to provide approximately an order-of-magnitude increase in FRB discovery
rate.

Although the field of view of CHORD is smaller than that of CHIME,
the improved sensitivity allows CHORD to probe deeper
into the Universe and detect intrinsically dimmer transient populations.
The broader frequency coverage extending to 1500\,MHz additionally
provides important propagation advantages, including reduced temporal
scattering and scintillation, together with improved preservation of
burst morphology and coherence for plasma and gravitational lensing
studies~\cite{CordesChatterjee2019}.

The outrigger stations located at HCO and GBO provide long baselines
for precise localization of transient events and VLBI follow-up,
allowing host galaxy identification and improved characterization of
burst environments.

\subsection{21 cm Intensity Mapping Cosmology}

One of the primary science goals of CHORD is precision measurement of
large-scale structure using 21\,cm intensity mapping across a broad
redshift range~\cite{CHORDWhitepaper2019}.
A central lesson from CHIME was that the dominant challenge for
21\,cm cosmology is not raw sensitivity alone, but precise control of
instrumental systematics and foreground contamination. Bright Galactic
and extragalactic synchrotron foregrounds exceed the cosmological
21\,cm signal by many orders of magnitude, requiring exquisite
instrumental stability and calibration accuracy~\cite{Shaw2014}. 

These considerations strongly shaped the CHORD design philosophy.
The telescope emphasizes highly redundant baseline distributions,
precision reflector manufacturing, and stable analog signal paths to improve calibration fidelity and suppress
frequency-dependent instrumental structure. The wide instantaneous
bandwidth additionally supports improved foreground characterization
and broad redshift coverage for baryon acoustic oscillation (BAO)
measurements and large-scale structure surveys.

\subsection{Spectral line Galaxy Surveys}

CHORD’s design will also enable widefield Galactic and extragalactic spectral line surveys. 
The combination of CHORD’s high sensitivity and wide band will enable the detection of individual HI-emitting galaxies over a broad redshift range. 
These observations require implementing higher spectral resolution than standard intensity mapping modes and motivate the inclusion of a dedicated high-resolution up-channelization mode within the digital backend. 
Current forecasts assume a spectral resolution of 23.7 kHz (a factor 8$\times$ increase from the 195.3 kHz native resolution of the CHORD F-engine), but a wide range of up-channelization factors are feasible across the band. Forecasts by Bij et al\cite{Bij2026} imply that widefield CHORD HI surveys will be particularly powerful for measuring the low-mass end of the gas-rich galaxy population and searching for gas-rich starless halos predicted by cosmology, and also to characterize the massive galaxy population at redshifts of approximately z=0.4.   

The spectral line backend is also well-suited to studies of hydrogen and helium radio recombination lines (RRLs) in the Milky Way and nearby galaxies. With 115 RRLs (H164$\alpha$ at 1477.3 MHz through H279$\alpha$ at 301.2 MHz) in the CHORD band, we expect to be able to stack approximately 70 RRLs after accounting for RFI. This will bridge the gap in sensitivity between existing RRL surveys~\cite{Liu2019SIGGMA,Anderson2021SPICY}
and optical H$\alpha$ surveys~\cite{Haffner2003WHAM}, probing the energy transport between bright H\,\textsc{ii} regions and the diffuse warm ionized medium.

\subsection{Pulsar Science and VLBI}

CHORD additionally supports dedicated pulsar search and beamforming
modes. The combination of large collecting area, wide bandwidth,
low system temperature, and flexible digital beamforming provides
strong sensitivity to 
millisecond pulsars, canonical pulsars, and long-period radio transients.
Dedicated tied-array beams will support pulsar searches, pulsar timing,
transient follow-up, and VLBI applications.



\section{Telescope Architecture}
\label{sec:telescope}





CHORD is a wideband radio interferometer based on a highly redundant large-$N$, small-diameter drift-scan array architecture optimized for survey science, precision sky mapping, and time-domain astronomy.
The telescope is designed to maximize survey speed and observing efficiency through continuous drift-scan operation while maintaining excellent control of instrumental systematics for precision 21\,cm cosmology measurements.

The CHORD core array is located at the Dominion Radio Astrophysical Observatory (DRAO) in British Columbia, Canada, and consists of 512 fixed 6\,m parabolic reflector antennas arranged in a compact, highly redundant configuration. 
Two additional outrigger stations, each containing 64 dishes, are being constructed at the Hat Creek Radio Observatory (HCO) in California and the Green Bank Observatory (GBO) in West Virginia. 
These outriggers provide long baselines for improved transient localization, VLBI capability, and higher angular resolution.

The telescope operates as a transit instrument with no motorized tracking system. 
Dishes are fixed in azimuth and observe the sky through Earth rotation, substantially simplifying the mechanical design and improving long-term pointing stability. 
The elevation angle of each dish can be adjusted manually to select the observing declination range, but this adjustment is infrequent and time consuming. 
Elevation repointings are expected to take place on a cadence of a few months or more, depending on the science prioritization and the achieved calibration efficiency.

CHORD operates over an instantaneous frequency range of  300--1500\,MHz using ultra-wideband feeds mounted at the prime focus of each reflector. 
Low-noise amplifiers integrated directly onto the feed assembly provide a receiver noise temperature of approximately 30\,K. Analog signals are transported over 125~m coaxial cable runs to nearby receiver huts, where they are digitized and processed by the digital backend.

The system has been designed to support very high aggregate data throughput while maintaining operational flexibility for commensal science observations.
The digital signal processing system is based on a scalable FX correlator architecture designed to support multiple simultaneous observing modes. These include a full $N^2$ correlator for precision imaging and calibration, static formed-beam transient search pipelines for FRB discovery, tracking beamformers for pulsar and VLBI observations, and high resolution processing modes for spectral line galaxy surveys.


\begin{table*}[t]
\centering
\caption{Summary of primary CHORD telescope parameters. Several values
remain subject to final optimization during commissioning.}
\label{tab:chord_parameters}
\begin{tabular}{lcc}
\hline
Parameter & Value \\
\hline
Core array location & \multicolumn{2}{l}{DRAO, British Columbia, Canada} \\
Core Dish Geometry & \multicolumn{2}{l}{D=6~m composite, f/D = 0.21} \\
Core array size & \multicolumn{2}{l}{512 $\times$ 6~m composite dishes (14,500 m$^2$ collecting area)}\\
Outrigger arrays & \multicolumn{2}{l}{two 64 $\times$ 6~m dish arrays} \\
Outrigger locations & \multicolumn{2}{l}{HCO and GBO} \\
Observing mode & \multicolumn{2}{l}{Drift scan with manual elevation adjustment} \\
Accessible declination range & \multicolumn{2}{l}{20-80 degrees} \\
Correlator architecture & \multicolumn{2}{l}{FX correlator} \\
Receiver noise temperature & \multicolumn{2}{l}{$\sim$30\,K} \\
Frequency range & 300\,MHz & 1500\,MHz \\
Field of view & 12 deg & 2.5 deg \\
Formed Beam Size & $17\times23$ arcmin & $3.4\times4.5$ arcmin \\
21~cm Redshift & 3.7 & $-0.05$  \\
\hline
\end{tabular}
\end{table*}

\section{Dish Design and Performance}
\label{sec:dishes}


Each CHORD antenna uses a precision 6\,m composite dish with $f/D \approx 0.21$, placing the focal plane slightly below the plane of the dish rim. 
This unusually deep reflector geometry reduces spillover and minimizes direct crosstalk between neighboring feeds. 
The deep-dish architecture additionally improves control of beam systematics, which is particularly important for foreground subtraction in 21\,cm intensity mapping observations.

The CHORD dishes were first prototyped in the D3A and D3A6 prototype programs \cite{Islam2020D3A,Islam2022D3A6}, which established the fabrication, alignment, and metrology techniques used for the production telescope.
Since redundant calibration techniques rely on nearly identical primary beams across the array, mechanical repeatability and stable beam performance were major design drivers throughout the reflector-development program. CHORD calibration targeted RMS surface precision averaging $\lambda/500$ at the center of the band (666 $\mu$m RMS).

\subsection{Fabrication}

The CHORD reflectors are fabricated as single-piece composite fiberglass structures using vacuum infusion techniques. 
Because the CHORD reflectors are fabricated as single-piece structures, transportation logistics strongly influenced the manufacturing strategy. 
The completed dishes are too large for economical long-distance transport, requiring construction of a dedicated fabrication facility near the DRAO site.
This approach simplified deployment logistics while enabling tighter quality control during reflector production.

The dishes are installed on screw-pile foundations, selected both for cost efficiency and to minimize environmental disturbance during construction.
This approach avoids extensive excavation and reduced the impact of deployment activities on the observatory site.

\begin{figure*}[t]
\centering
\begin{minipage}{0.49\textwidth}
    \centering
    \includegraphics[width=\linewidth]{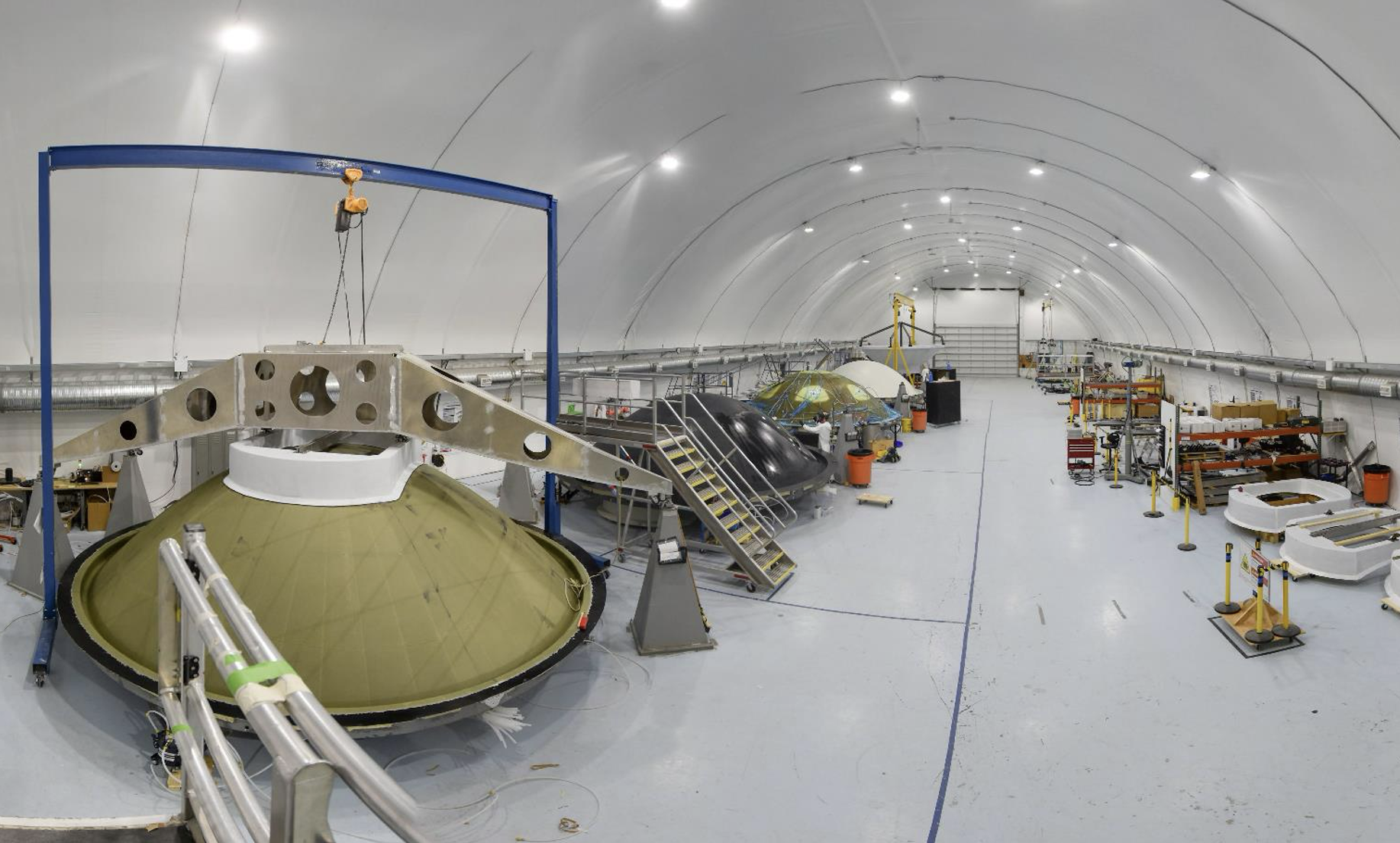}
\end{minipage}
\hfill
\begin{minipage}{0.49\textwidth}
    \centering
    \includegraphics[width=\linewidth]{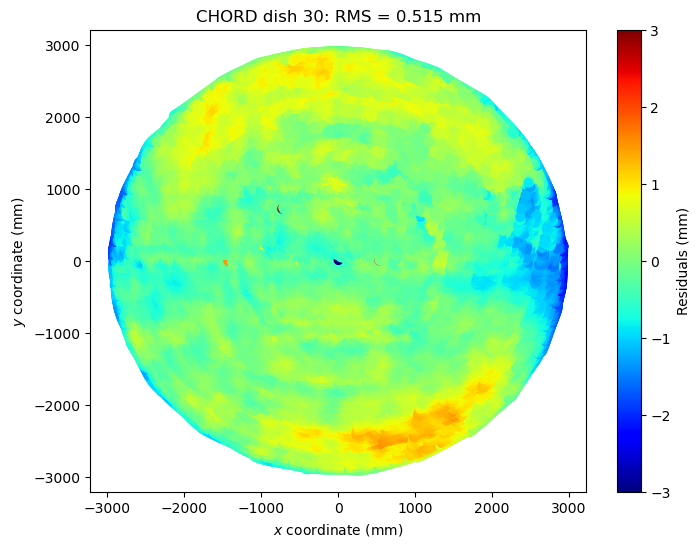}
\end{minipage}

\caption{ 
Left: Production of CHORD composite reflectors at the dedicated fabrication facility constructed near the DRAO site. The one-piece reflector design minimizes assembly complexity and improves dish-to-dish repeatability across the array.
Right: Laser metrology measurements of a typical CHORD reflector surface. Extensive metrology studies were used to validate the composite fabrication process and quantify reflector surface precision and alignment repeatability. 
The fabrication process achieves a typical surface precision of 
$550~\mu$m RMS, supporting the stable and repeatable beam performance required for precision redundant calibration.
}
\label{fig:dish_construction_metrology}
\end{figure*}

\subsection{Manufacturing Precision and Metrology}

Extensive laser-tracker and photogrammetry measurements were performed during the prototype development program to characterize reflector surface precision, feed alignment repeatability, and mechanical stability 
\cite{Islam2022D3A6,Islam2026PerformanceModeling,Islam2026CompositeReflectors}.
These measurements demonstrated that the composite
fabrication process can achieve the precision required for highly redundant 21\,cm intensity mapping arrays.

All dishes are produced from a set of four precision molds engineered to be nearly identical.  
The achieved mold surface precision of 250 $\mu$m RMS required extensive iterative metrology and many weeks of manual sanding and surface correction during mold preparation.


For the production array, surface characterization was performed for the first 35 dishes as they came off the mold. The measured precision of the dish surface with respect to the intended paraboloid is $550~\mu$m RMS with a total spread of 350--700 $\mu$m RMS, exceeding the project target of 666 $\mu$m$_{RMS}$.
For redundant calibration, the difference between dish surfaces is of great importance. 
The data from the first 35 dishes were gridded, clocked, stacked and averaged so that point to point analysis can be performed and the average surface measured.
The measured precision of the dish surface with respect to the average of the dish surfaces is $294~\mu$m RMS.
For the remainder of the array dish production, surface characterization will be performed for every 10th dish to check for any degradation of manufacturing.





\section{Feeds and analog system}
\label{sec:Feeds}



\begin{figure*}[t]
\centering
\begin{minipage}{0.38\textwidth}
    \centering
    \includegraphics[width=\linewidth]{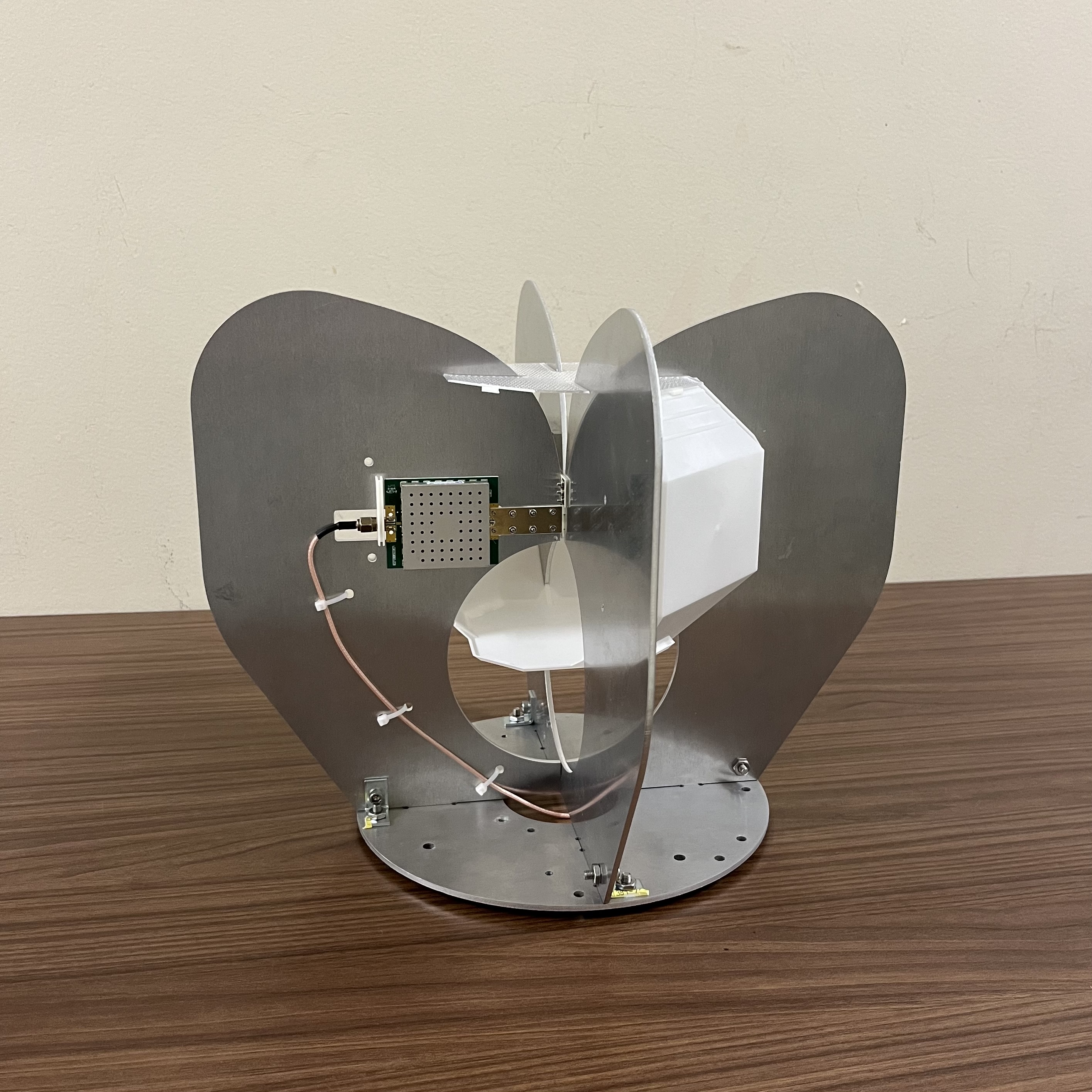}
\end{minipage}
\hfill
\begin{minipage}{0.58\textwidth}
    \centering
    \includegraphics[width=\linewidth]{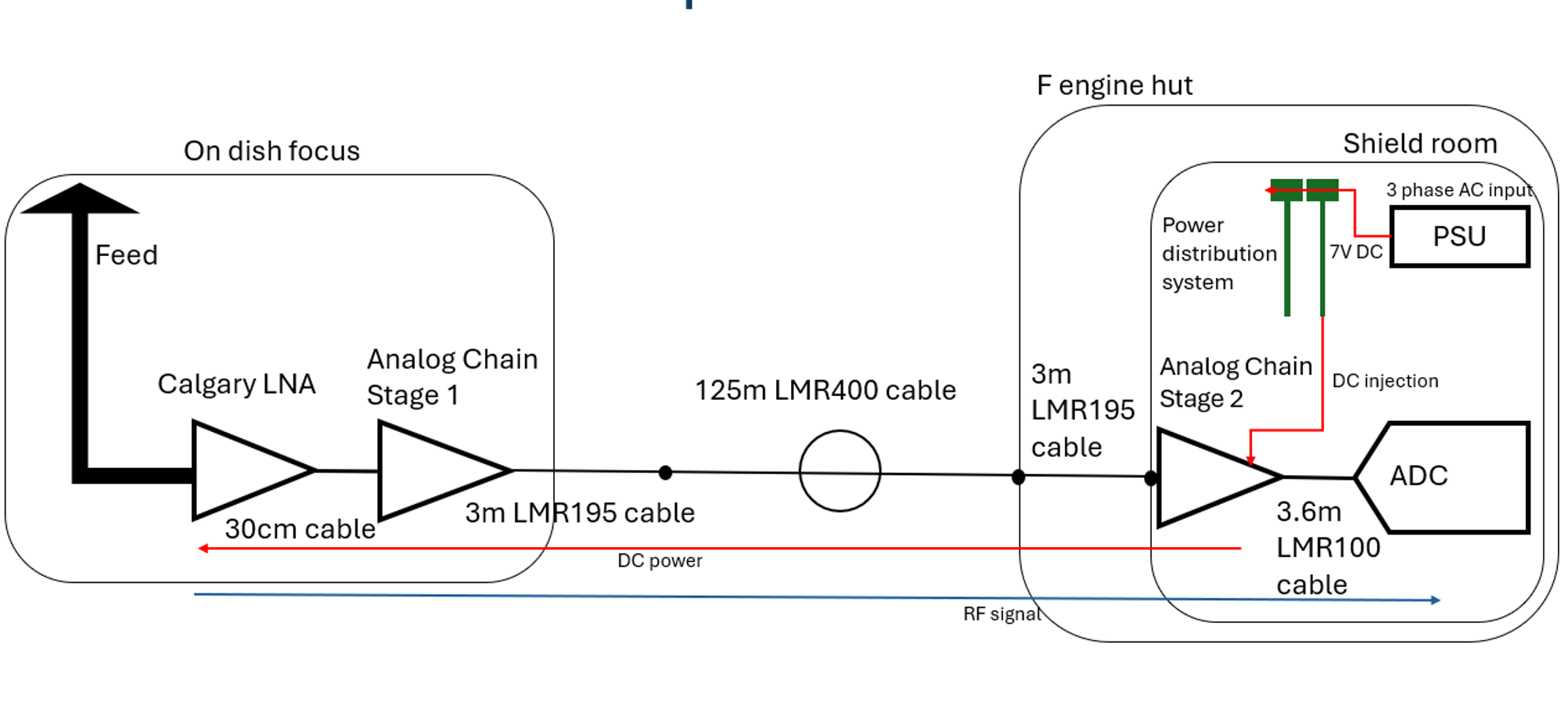}
\end{minipage}
\vspace{4pt}
\caption{
Left:
Prototype CHORD ultra-wideband feed assembly, including the dual-polarization feed structure, integrated balun, and low noise amplifier. 
A section of the white support enclosure has been removed to expose the receiver electronics for clarity. The compact geometry minimizes blockage on the deep six-meter $f/D\simeq $0.21 CHORD reflectors while enabling wideband performance across the 300--1500\,MHz observing band.
The first-stage low-noise amplifiers are integrated directly at the feed terminals to minimize pre-amplifier losses. \\
Right: Simplified analog signal chain. Signals are amplified at the feed, transported via coaxial cables to the receiver enclosure, conditioned by additional amplification and filtering stages, and subsequently digitized by the 
RFSoC-based
F-engine system.
The coaxial cables also carry DC power from the F-engine to the electronics at the feed.
}
\label{fig:CHORDFeed}
\end{figure*}

Each CHORD dish is equipped with a compact dual-polarization ultra-wideband feed covering the 300--1500\,MHz observing band, shown in Figure~\ref{fig:CHORDFeed}(left). The feed design was developed specifically for large-$N$, small-diameter interferometric arrays, where low cost, manufacturability, and compact physical size are critical design constraints. The architecture is based on a modified Vivaldi-style feed with an oversized backshort, allowing broad fractional bandwidth while maintaining a compact geometry appropriate for the deep ($f/D \approx 0.21$) CHORD reflectors \cite{MacKay2023Feed}.  
The feed structure is fabricated from CNC machined aluminum petals with integrated PCB baluns and support structures, enabling inexpensive and repeatable large-scale production. 
After prototyping with laser-cut petals, CNC machining was adopted to achieve the necessary tolerance, since feed asymmetry affects beam shape and performance.

The CHORD analog low noise amplification~\cite{10124757}, filtering, and signal transport chain is shown in Figure~\ref{fig:CHORDFeed}(right).
Low-noise amplification is integrated directly at the feed terminals in order to minimize pre-amplifier loss. The first-stage amplifier uses a custom wideband PCB design together with commercially available low-noise components, and is optimized jointly with the feed impedance to achieve low receiver temperature across the full observing band. Direct attachment of the amplifier to the balun structure avoids connector losses ahead of the first gain stage, which is particularly important for maintaining low system temperature at the low-frequency end of the band.

The analog signal chain is intentionally simple and robust. After the feed and first-stage amplification, signals are transported over 125~m coaxial cabling to the receiver enclosure, where additional gain, filtering, and conditioning stages are applied prior to digitization.  DC power is delivered to the feed electronics by the same coaxial cable.  The design philosophy emphasizes stability, repeatability, and scalable deployment across the full CHORD array, while minimizing analog complexity ahead of the digital backend. Careful gain distribution and band-defining filters suppress out-of-band interference while maintaining stable operating margins throughout the chain.

The feed and receiver system operate entirely at ambient temperature. Simulations and laboratory measurements of the prototype system indicate low material losses, good impedance matching across the observing band, and receiver temperatures consistent with the CHORD system requirements \cite{MacKay2023Feed}. Detailed on-sky performance measurements are presented in Sections~\ref{sec:performancethreedish} and \ref{sec:performance}.

\section{F-Engine}
\label{sec:Fengine}



 




The CHORD core-array F-engine consists of 128 t0.technology Control and Readout System (CRS) FPGA boards~\cite{Montgomery2024CRS} based on the Xilinx Zynq Ultrascale+ Radio Frequency System on a Chip (RFSoC) platform.
The system digitizes signals from 512 dual-polarization feeds, corresponding to 1024 analog inputs total. Each CRS board processes eight analog inputs and directly samples the full CHORD observing band at 3.2\,GSPS with 14-bit precision.

The original CHORD concept proposed reuse of the ICE digitization platform developed for CHIME~\cite{Bandura2016}. 
Applying the ICE architecture to the substantially broader CHORD bandwidth would have required division of the observing band into multiple analog sub-bands with independent digitization chains.
In the intervening years, the CRS RFSoC platform was developed, enabling direct digitization of the full 300--1500\,MHz CHORD band within a single frontend system.
This architecture substantially simplifies wideband calibration and provides improved control of gain stability and frequency-dependent instrumental systematics across the observing band.

Hendricksen\cite{HendricksonThesis} performed a detailed comparison of the CRS RFSoC platform against the ICE-based architecture originally proposed for CHORD, including measurements of nonlinearity, effective number of bits (ENOB), spurious-free dynamic range (FDR), and wideband spectral performance. 
The noise spectral density (NSD) of the CRS RFSoC was measured to be $-154$~dBFS/Hz (compared to $\approx $-132.6~dBFS/Hz for the ICE system), corresponding to an ENOB of 10~bits (6.9--7.2~bits for ICE), with the CRS RFSoC adding 3 effective bits of dynamic range. The SFDR of the CRS RFSoC was measured to be $< -73.3$~dBFS (compared to $< -49$~dBFS for the ICE system), which includes contributions from spurious tones caused by the interleaving of sub-ADC tiles, demonstrating that these products are suppressed to a level much below the largest spurious tones of the ICE ADC. Finally, the crosstalk of the CRS RFSoC was measured to be $< -72$~dBFS for adjacent channels, while the ICE ADC crosstalk is only $< -46$~dBFS for channels on the same chip, and $< -66$~dBFS for channels on separate ADC daughter mezzanines, demonstrating a substantial improvement in leakage from neighboring channels.
These studies demonstrated that the CRS system achieves performance comparable to or exceeding that of the ICE architecture across all measured metrics while simultaneously eliminating the need for analog sub-band splitting which would have introduced the potential for substantial systematic errors in complex gain versus frequency.

Following digitization, data are channelized on-board using a resource-efficient CASPER-style~\cite{CASPER} polyphase filter bank (PFB) followed by a 16384-point FFT.
The resulting 8192 frequency channels have a bandwidth of 195.3125\,kHz per channel. Following channelization and requantization (described below), the data undergo an initial on-board corner-turn operation that reorganizes the eight locally digitized inputs prior to transmission over 100\,GbE links to the downstream networking fabric.

The current CHORD implementation performs the remaining corner-turn operations entirely through an Ethernet switch network before delivery to the X-engine processing systems. The CRS architecture was designed to support additional corner-turn modes, including backplane-level and inter-crate corner-turn operations, providing future flexibility for alternative correlator and networking configurations~\cite{HendricksonThesis}.

Prior to requantization, a programmable complex digital gain is applied independently to each frequency channel and input in order to equalize signal levels across the observing band and optimize quantization efficiency.
This stage additionally provides operational flexibility for mitigating strong narrowband radio-frequency interference (RFI) and compensating for frequency-dependent analog gain variations before requantization to complex $(4+4i)$-bit precision.

\begin{table*}[t]
\centering
\caption{CHORD FX-correlator Parameters.}
\label{tab:fengine}
\begin{tabular}{ll}
\hline
Parameter & Value \\
\hline
F/X-engine architecture & FPGA-based F-engine / GPU-based Correlator \\
F-engine platform & t0.technology CRS RFSoC \\
FPGA device & AMD Zynq Ultrascale+ RFSoC \\
Number of CRS boards & 128 in eight crates \\
Analog inputs per board & 8 \\
Total analog inputs & 1024 \\
ADC sample rate/bits & 3.2\,GSPS at 14~bits \\
FFT length & 16384 \\
PFB/FFT channels and bandwidth & 8192 channels at 195.3125 kHz \\
PFB implementation & 4-tap CASPER-style PFB \\
Output requantization & $(4+4i)$ bits \\
Output interface & 100\,GbE \\
Corner-turn implementation & On-board + Ethernet switch \\
Aggregate front-end digitized data rate & 1024 inputs at 3.2 GSPS, 14 bits = 46 Tbps \\
F-X Corner turn network rate & 1024 inputs $\times$ 6144 frequency channels $\times$ $(4+4i)$ bits = 9.83 Tbps \\ 
GPU host platform & 64x Dell PowerEdge R760XA \\
CPU number and type & $128\times$ Intel Xeon 5416S (16-core) \\
GPU number and type & $128\times$ Nvidia A40 \\
GPU host network & $2 \times$ 100\,GbE Input NICs, $2\times (2\times25)$\,GbE Output NICs \\
System RAM & 2 TB per node, 128 TB total. \\
$N^2$ Correlation integration time & Variable 1--30\,s with fringe stopping \\
Number of formed beams & $5000\times$ 4-bit intensity beams $\times$ 32K frequencies ($\sim$1.3 ms time resolution) \\
Intensity beam data rate & $\sim$500\,Gb/s \\
Baseband call-back buffer depth & $\sim$100\,Seconds \\
Coherent beam data rate & 9.6\,Gb/s/beam (up to 96 beams) \\
\hline
\end{tabular}
\end{table*}

\begin{figure}[t]
\centering
\includegraphics[width=0.5\textwidth]{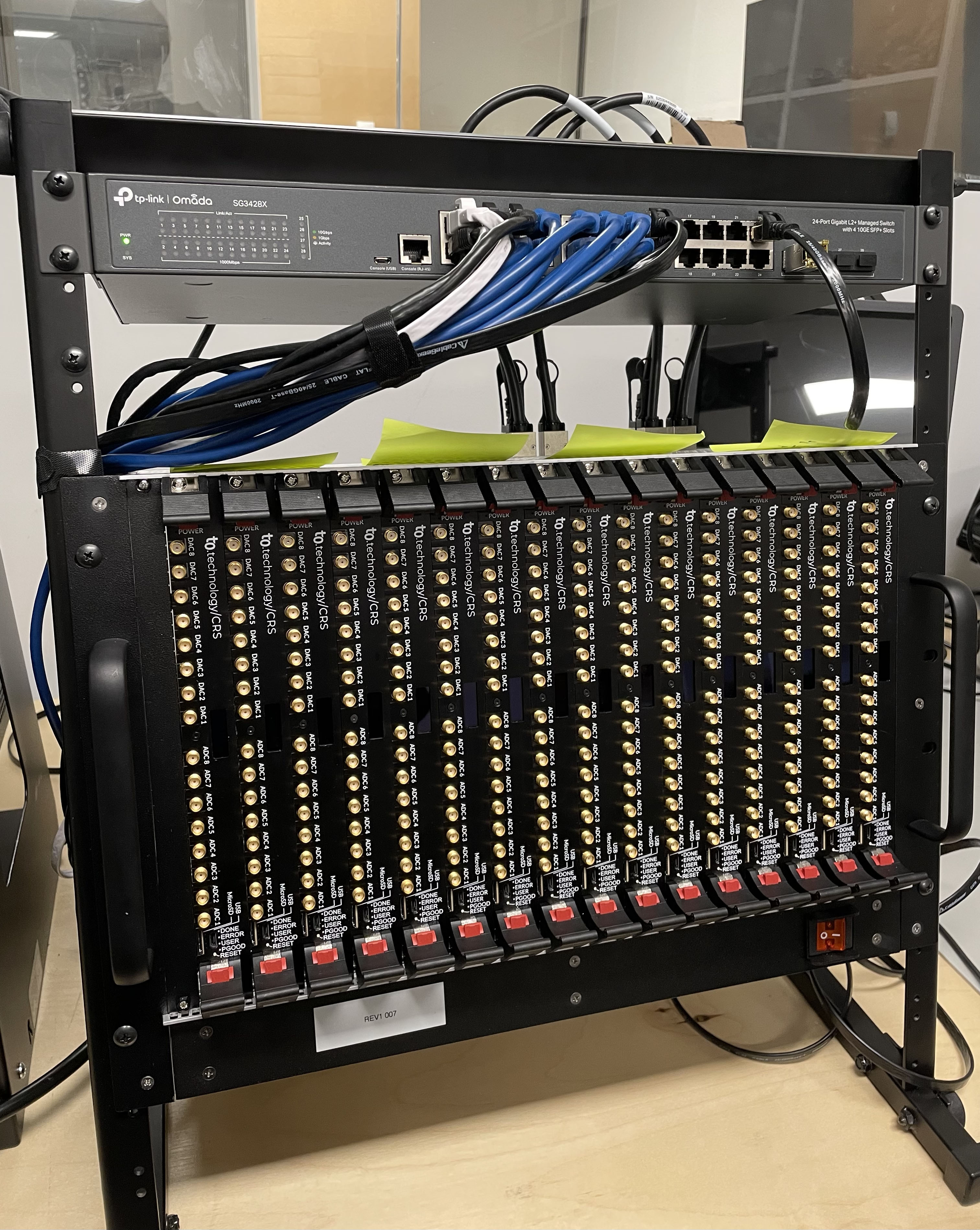}
\caption{
CRS RFSoC F-engine hardware used for digitization and channelization
of the CHORD frontend signals. The photo shows one crate housing 16 CRS boards. The CHORD core array will use a total of 8 crates (128 CRS boards).
}
\label{fig:crs_crate}
\end{figure}

\subsection{F-engine Firmware and Software}
\label{sec:fengine_firmware}





\begin{figure*}[t]
\centering
\includegraphics[width=\textwidth]{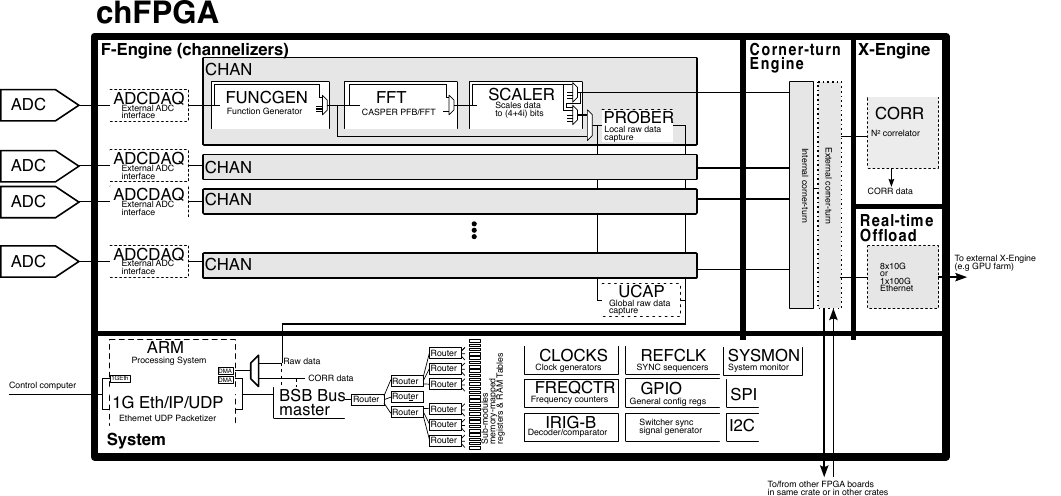}
\caption{
Simplified block diagram of the CHORD F-engine firmware architecture. 
Digitized signals from the analog front-end are channelized and processed within RFSoC-based FPGA boards before transmission to downstream corner-turn and X-engine systems. The firmware supports realtime channelization, packetization, synchronization, monitoring, and multiple simultaneous observing modes.
}
\label{fig:fengine_firmware}
\end{figure*}

The CRS F-engine signal path described above is implement in the \texttt{chFPGA} firmware framework described in Ref~\cite{Hendricksen2026PocketCorrelator}. \texttt{chFPGA} is a flexible open-source signal-processing platform developed by McGill University and t0.technology for wideband radio interferometer front-end systems~\cite{chFPGA}. Figure \ref{fig:fengine_firmware} illustrates the internal architecture of the firmware. It supports realtime digitization, a bank of channelizers that include embedded test signal generators and statistics engines, two switchable high-dynamic-range post-FFT floating point gains tables, $(4+4i)$-bit quantizer, independently configurable science and monitoring outputs, a single-board or multi-board corner-turning engine, multi-board synchronization logic, a fully programmable switch- and GPU-friendly data packetizer/100G Ethernet transmitter, and an optional FPGA-based X-Engine for single-board correlator operations. The monitoring outputs allow full-resolution FFT data captures that facilitate fast gain computations. Although originally developed for CHIME, the firmware can now be implemented for multiple FPGA platforms and for multiple configurations, including the CHORD firmware running on the CRS platform. For large arrays, the firmware uses the crate’s shared backplane to implement deterministic, sub-sample multi-board synchronization and data corner-turning.

System operation and monitoring are controlled through the \texttt{pychfpga} Python software package, which reads a configuration file, finds the target boards on the network and connects to them, programs the on-board ARM and FPGA firmwares, configures the FPGA firmware for the selected signal processing functions, and performs the array-wide synchronization. Once data acquisition is started, an HTTP/REST server provides a real-time control and monitoring interface that can be used to change the system configuration during observatory operations, and continuously publish real-time status information on every part of the hardware and firmware that can be gathered and displayed with Prometheus\cite{Turnbull2018Prometheus} and Grafana\cite{Grafana} systems monitoring and visualiozation tools. A separate data acquisition server gathers the periodically captured ADC, FFT and optionally X-Engine data, serves real-time statistics, and saves the data to disk.


\section{X-Engine and Networking}
\label{sec:xengine}

\begin{figure}[htbp]
    \centering
    \includegraphics[width=1.0\textwidth]{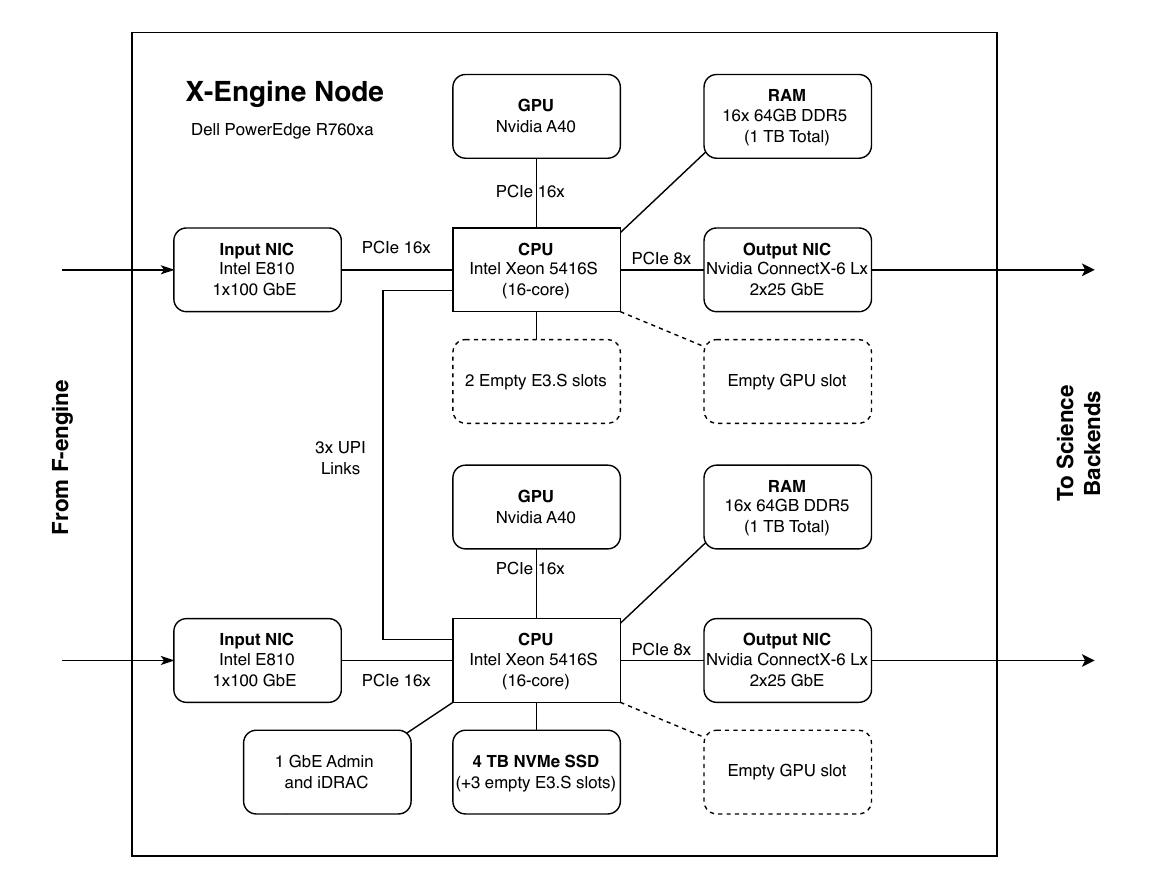}
    \caption{Block diagram showing the main components of the X-engine nodes.  The main item driving the design is the amount of RAM required for the baseband buffer.  Because 64 GB DIMMs are cheaper than larger DIMMs on a cost/GB basis, this required a system with a large number of DIMM slots (32 total between the two CPUs). The systems have 2x unused GPU slots, and 5 unused PCIe 5.0 NVMe slots with an E3.S form factor, which could be used for future upgrades.}
    \label{fig:x-engine-hardware}
\end{figure}

\begin{figure}[htbp]
    \centering
    \includegraphics[width=0.76\textwidth]{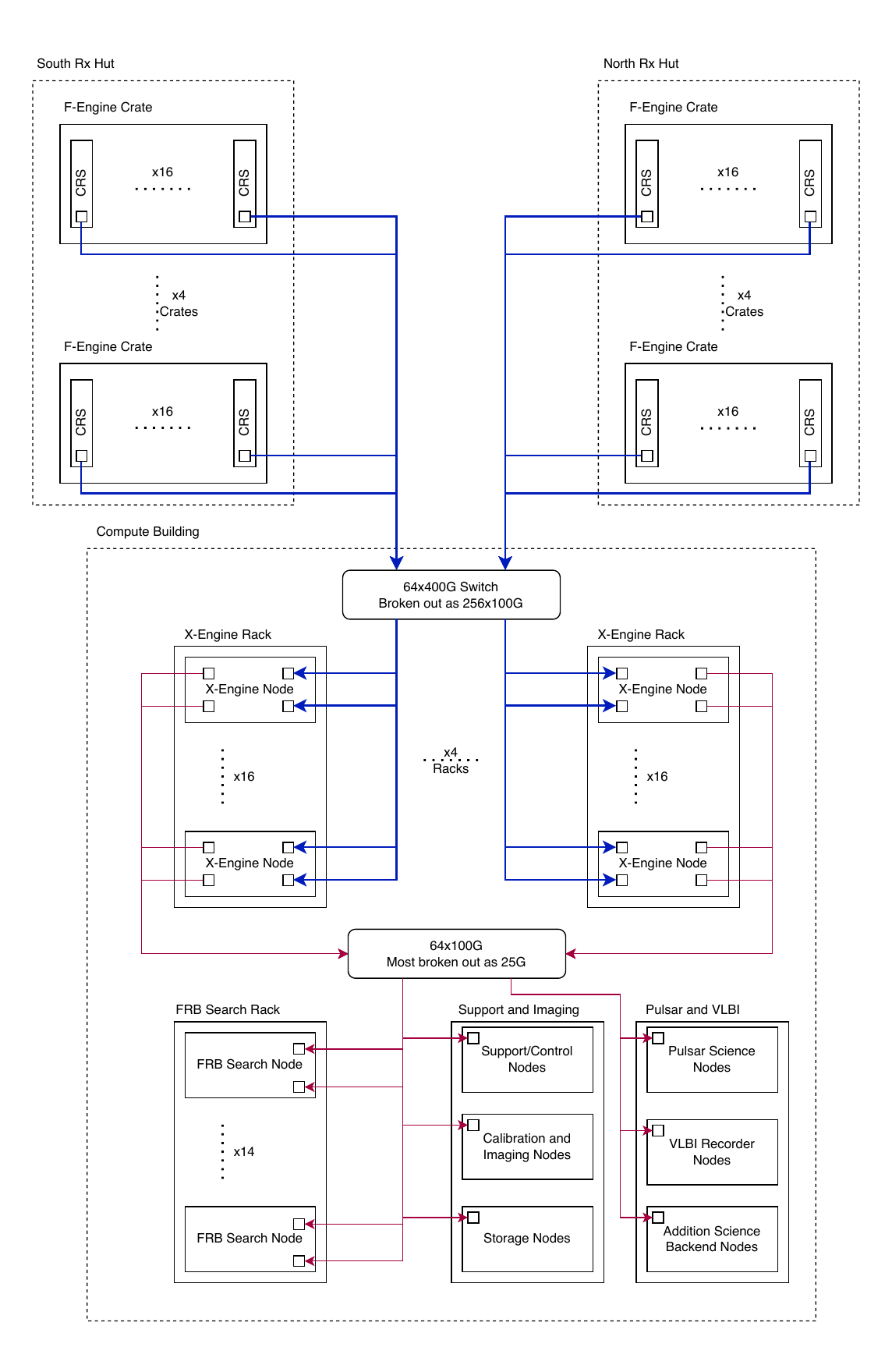}
    \caption{High-level CHORD networking and X-engine architecture. Frequency-domain
data generated by the CRS RFSoC F-engine systems are packetized and
transported over a high-speed Ethernet switching fabric that performs
the global corner-turn operation prior to GPU-based X-engine processing.
The resulting correlated and beamformed data products are distributed to
multiple simultaneous realtime science backends including imaging,
calibration, FRB search, pulsar timing, and VLBI systems.}
    \label{fig:chord_network}
\end{figure}

The CHORD X-engine is built around commercial-off-the-shelf (COTS) GPU compute servers, running highly optimized real-time processing software presented in Section \ref{sec:pipeline}.   The X-engine is comprised of 64 Dell PowerEdge R760XA rack-mount servers, each with 2 CPUs, 2 GPU, 4 NICs, and 2 TB of RAM.  See Figure \ref{fig:x-engine-hardware} for a block diagram of the X-Engine hardware and its interconnects, and also Table \ref{tab:fengine} for details of the hardware. Data from the F-engine is sent to the X-engine via a 64-port 400 GbE switch, with each port broken out to $4\times100$ GbE links.  Each X-engine node connects to this switch with two 100 GbE links, and sends data to the science backends via two 25G GbE links, shown in Figure \ref{fig:chord_network}. A COTS server system with GPUs and Ethernet networking was chosen for CHORD to maximize the flexibility to add new science use cases, and allow for rapid changes to the system. 

A full mesh transpose of the data is required in both the F-engine to X-engine, and X-engine to science backend networks.  Each of the 128 boards in the  F-engine outputs data for 6144 frequency channels, 8 dishes and 1 polarization, while the X-engine processes only 48 frequency channels per GPU, but requires data from all 512 dishes and both polarizations.  This transpose is accomplished by having the F-Engine firmware internally assemble the board's data into 128 packets, each packet with 48 frequency channels and a unique destination IP and MAC address. After the network switch routes the packets, each of the two NICs on each X-engine node receives one of those packets from each CRS board.  At the nominal digitization rate of 3.2 GS/s, 4+4-bit complex encoding, and down selecting to 1200 MHz of bandwidth (6144 frequency channels) results in an aggregate data rate through the switch of 9.8 Tb/s, and 153.6 Gb/s/node, or 76.8 Gb/s/GPU (plus some small header overhead). 

Once on the GPU, the electric field data is processed into a number of different products depending on the backend requirements of that science case.   These are described in more detail in Section \ref{sec:pipeline}, see also Figure \ref{fig:x-engine-dataflow} for a block diagram of the data flow.   All of these data products are then transferred off the GPU back into the CPU's DRAM, where they are sent out to the different science backends using the 25 GbE network links. Like the F$\to$X network, many of these backends require another full transpose of the data.  For example, the FRB pipeline produces all beams, but only a subset of frequencies per GPU, however the search system requires a subset of beams and all frequencies.  To achieve this, a second switch will be used with $64\times100$ GbE ports, which can each be broken out into $4\times25$ GbE links, for a total of $256 \times 25$ GbE ports.  We will initially connect only two of the four 25 GbE links per node (one per CPU), as this is all we require for the initial science goals.  However, we maintain the option to use a second switch to connect the other two 25 GbE links for future science cases that require additional backends.

\section{Realtime Software Pipeline and Science Backends}
\label{sec:pipeline}

\subsection{Overall design}

\begin{figure}[htbp]
    \centering
    \includegraphics[width=1.0\textwidth]{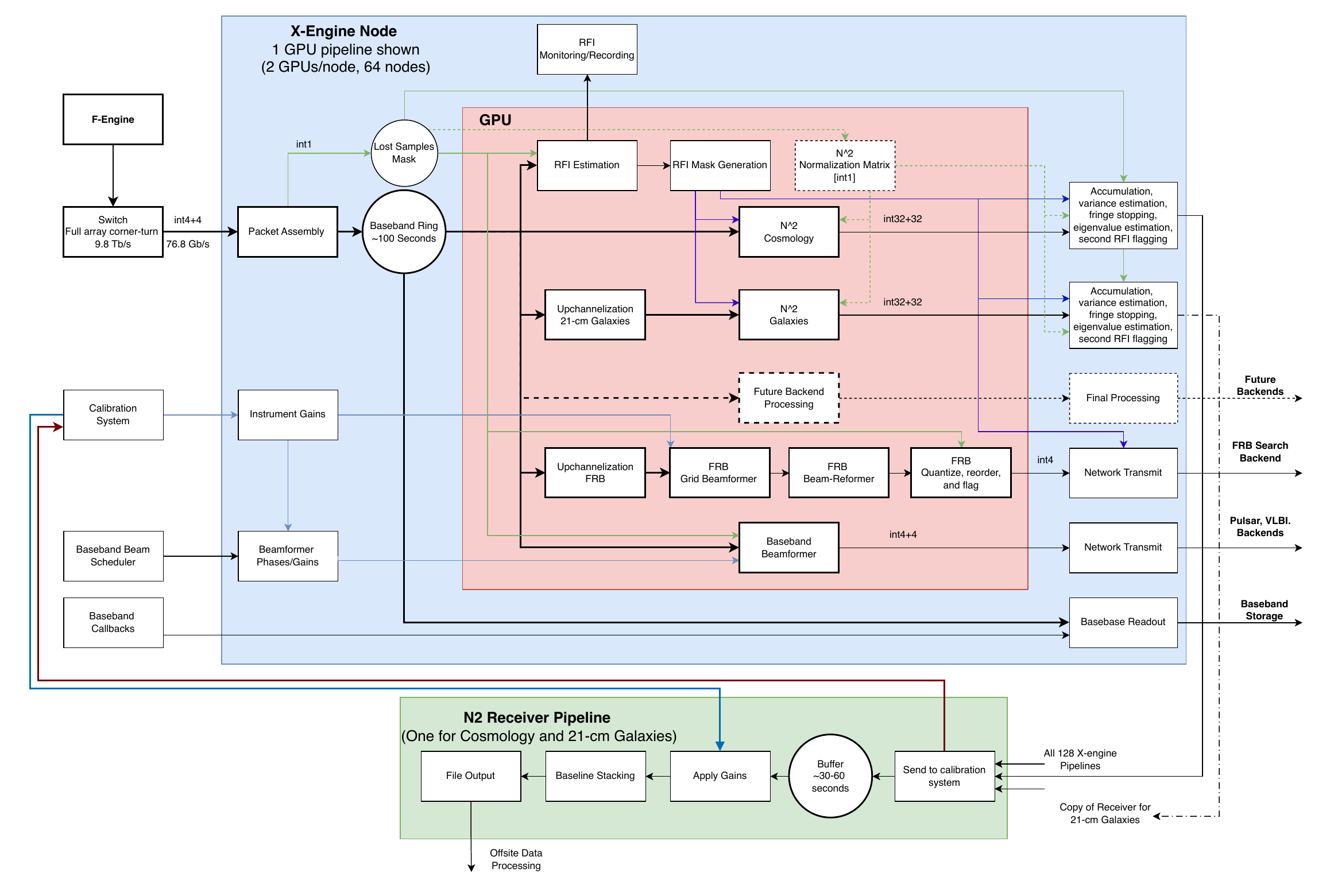}
    \caption{Data flow through the X-engine.  Data arrives from the F-engine via the corner-turn switch, and is assembled into an array of baseband voltage timestreams, which is buffered in CPU DRAM for up to 100 seconds.  A copy of this data is transferred immediately to the GPU, where the majority of the data processing takes place.  Along with the baseband data, an array of packet-loss information is transferred to the GPU to allow for renormalization and/or flagging of missing data.     Note that only one GPU path is shown; there are 2 GPUs in each node running a copy of this pipeline, and 64 nodes in total.  The $N^2$ packet-loss normalization matrix is listed as an optional feature.  There are two ways to handle packet loss in the corner-turn, either flag all samples if one part of the array is lost, or track which parts of the array are lost, and generate a normalization matrix.   The latter retains more data, but introduces a lot of complexity for the down stream analysis; the final design has not been decided.    
    }
    \label{fig:x-engine-dataflow}
\end{figure}


The real-time software pipeline lives on the X-Engine nodes. An
individual X-Engine node's input are the corner-turned voltages, i.e.
timestreams containing voltage measurements from 512 antennas, each with two polarizations for a subset of
frequency channels. The overall architecture splits the data streams
by frequency channel and each channel is processed independently. Each
channel is processed by various stages in a dataflow pattern where later stages consume the output of earlier stages through buffers, and all stages run quasi-simultaneously. The majority of
science processing tasks ($N^2$ correlation, calibration, FRB search,
pulsar beamforming) are described below and shown in Figure \ref{fig:x-engine-dataflow};
each task can correspond to tens of separate stages in the final pipelines.

Most of the data processing occurs on the 128 Nvidia A40 GPUs, with
networking code, buffering code, and a small fraction of the data
processing living on the CPUs. Each GPU handles 48 frequency channels
simultaneously. Input voltage data are generally processed in batches of $8192$
time samples or $41.943\,\mathrm{ms}$, except for some upchannelization kernels, which require different time windows.   This is possible thanks to ring-buffer logic in the GPU memory space, which allows reading different ranges of input data by different kernels, down to the byte level.  

Our software infrastructure \texttt{kotekan} \cite{kotekan} is an
updated, extended, and more flexible version of the infrastructure
used by CHIME. \texttt{Kotekan} provides a modular architecture where
\emph{stages} process data and communicate via \emph{buffers} holding
data. \texttt{Kotekan}'s architecture is highly parallel, both on CPU cores and
GPUs, manages the dataflow to ensure data consistency, and runs stages
commensurate with the amount of input data available. Both the amount
of data processed by a stage as well as the intervals at which a stage
runs are fully dynamic. This allows our dataflow graph to handle
different time downsampling factors in different parts of our
pipeline, as dictated by the differing time resolutions for different
science tasks.  An overview of the processing is included here, and a more detailed description will be published in a future paper. 

\subsection{Packet capture and corner-turn}

Packets are captured in the CPU using a custom wrapper on the Data Plane Development Kit (DPDK) for kernel bypass and interacting with the NIC.  The data is then transposed into a format which is optimal for processing on the GPU, and finally transferred to the GPU with a DMA from CPU DRAM to GPU DRAM.  A copy of this data is buffered in the CPU DRAM for about 100 seconds and provides a call back buffer for real-time detections.

\subsection{Baseband callback buffer}

One of the biggest drivers of the design of the X-Engine is optimizing the amount of RAM available for baseband buffering.  Since it is infeasible to search for FRBs and other short-duration transients at baseband data rates, the search must take place on a reduced set of data.  These are formed intensity beams in the case of CHORD (see Section \ref{ssec:frb_search}).   Once a transient event has been detected, it is extremely useful to have a copy of the baseband voltages from that event, as this can provide better localization, rotation measure data, enable VLBI with the outriggers, and more.  To make this possible we provide a callback API on the X-Engine which takes the time of the event, a dispersion measure (DM), and width, and then sends the corresponding baseband data to a storage machine on site.   The CHORD system has been designed around providing an approximately 100 second buffer, which after accounting for detection latency, should be able to record the voltages for events out to around DM 2000 pc/cc.

\subsection{RFI Flagging}

The GPU correlator computes a boolean real-time RFI mask that is applied when computing visibility matrices, following the strategy developed for CHIME\cite{Taylor_2019}. The masking happens in multiple stages on different timescales. We compute summary statistics (per-channel variance and spectral kurtosis (SK) statistic) on short timescales, and mask RFI on \mbox{$\sim$1 ms} timescales on the GPU. These summary statistics are sent to the CPU, where they are used to mask RFI on $\sim$30 ms timescales. Finally, the summary statistics are downsampled and written to the visibility data files, so that flexible ``offline'' RFI masking can be performed on longer timescales ($\gtrsim 1$ sec). When we compute the masked visibility matrix on the GPU, we also compute the covariance matrix of the RFI mask (using 1-bit tensor core Parallel Thread Execution (PTX) instructions), in order to correctly normalize.

\subsection{FRB search pipeline}
\label{ssec:frb_search}

The FRB beamforming and search pipelines are the most complex parts of the CHORD real-time system. The beamforming pipeline runs on the X-engine and creates a ``data cube'' of intensities indexed by (beam, frequency, time). The search pipeline runs on a dedicated backend, and searches each beam in parallel for FRBs, i.e. single pulses whose dispersion exeeds a Milky Way model. We briefly summarize the FRB beamforming and search pipelines here; details will be presented in a future paper.

On the X-engine, we first apply frequency-dependent upchannelization to the input voltages, in order to obtain frequency resolution which varies from 6--200 kHz at the bottom/top of the CHORD frequency band (300--1500 MHz). Then, we form a regular $48 \times 48$ spatial array of beams, and sinc-interpolate (in a separate Compute Unified Device Architecture (CUDA) kernel) to a specified set of sky locations. It can be shown (by an argument similar to the Nyquist sampling theorem) that this interpolation agrees to machine precision with exact beamforming at the target sky locations. The beams are quantized to 4 bits and sent over the network to the FRB search backend.

On the FRB search backend, the first step is building a real-time boolean RFI mask, to remove data which has been contaminated by RFI. Our approach to RFI masking builds on our approach for CHIME\cite{Rafiei-Ravandi:2022rwl}, but the low-level filtering steps are now performed on the GPU instead of CPU. Next, we apply a matched filtering search for FRBs (sometimes called the ``dedispersion transform''). Our GPU-based implementation of the search is heavily based on our CPU-based CHIME search, but is a factor $\sim$10$^2$ times more efficient, in order to process higher data rates in CHORD.

\subsection{Pulsar and VLBI pipeline}
The pulsar (baseband) beamformer is a single GPU kernel that forms coherent beams tracking a specified set of sky locations. For each frequency channel, polarization, and time sample, the input data is a length-512 {\tt int4+4} vector, where 512 is the number of dishes. We multiply by a $B$-by-512 {\tt int8+8} phase matrix, where $B$ is the number of formed beams. The output is a length-$B$ complex vector, which we quantize to {\tt int4+4} and write to GPU memory. 

The matrix multiplication is performed with a custom CUDA kernel which uses tensor core PTX instructions for speed, does not write intermediate arrays to global GPU memory, or read the same input data in multiple threadblocks. This custom kernel is so fast that we can form $\sim$10$^2$ baseband beams using a few percent of X-engine GPU compute. Since the phase matrix is small, it can be recomputed on the CPU every $\sim$100 ms, so implementing tracking beams is straightforward. Baseband beams can either be sent to a backend (e.g. for pulsar timing) or saved to disk (e.g. for offline VLBI).

\subsection{21 cm cosmology pipeline}

The real-time cosmology backend writes visibility matrices to disk with 195 kHZ frequency resolution and $\sim$30 second time resolution. This coarse time resolution is feasible because visibility matrices are ``fringe-stopped'' in order to remove the leading-order effects of sky rotation. The endpoints of the time bins are fixed in LST, in order to facilitate ``stacking'' over sidereal days in our offline analysis pipeline.

For 21-cm intensity mapping, visibility matrices are processed by a map-making pipeline which is similar to that used for CHIME \cite{Shaw:2014khi,CHIME:2025cee}, but adapted to the CHORD dish layout. A separate pipeline is under development for finding spatially unresolved 21-cm emitting (galaxies) or absorbing sources\cite{hans,Bij2026}. For the latter pipeline, we will also produce a separate visibility matrix data product which is upchannelized in frequency, and restricted in frequency range, in order to increase SNR for galaxies at low redshift ($z \lesssim 0.1$) with low velocity dispersion.

The full visibility matrices will be too large to store on disk for the multi-year duration of the CHORD project. We plan to store the full visibility matrices for a limited time window, which will suffice to assess data quality and obtain calibration solutions. For long-term storage, the data volume will be reduced by averaging redundant baselines after applying the calibration solution.

\subsection{Operations and monitoring}

The vast majority of electronic hardware and software components in the pipeline provide metrics that measure system health and throughput. These range from environmental and compute temperatures to system and event logs to operational throughput. These metrics are continuously scraped from all available sources using a system-wide Prometheus~\cite{Turnbull2018Prometheus} service, stored in full for a 30 day period, and a representative subset of this data is stored indefinitely. This data can be queried by Grafana\cite{Grafana} users to visualize the system status, while alert services use it to issue messages about urgent system issues.

Installation and maintenance of software on most systems is accomplished using a suite of \texttt{ansible} and baseboard management controller (BMC) scripts. These scripts provision CHORD services with BIOS settings, host OS installation, network and user management configuration, and software packages used by different CHORD services depending on the machine targeted for installation.

While data processed by real-time CHORD systems is anticipated to be on the order of 10 TB/s, only highly reduced data products are stored on disk at a rate closer to 1 TB/day. The bulk of this data consists of the occasional raw baseband voltage dumps, and visibility correlation matrices. However, data derivatives and associated metadata are also written, including the full system configuration during data collection, timing information, telescope geometry information, and more. Due to the large storage requirements, data is regularly transferred off site for further processing, analysis, and long-term storage.

  

\section{Infrastructure and RF Shielding}
\label{sec:housing}



The CHORD F- and X-engine systems are housed within
shielded equipment enclosures designed to minimize self-generated
RFI while providing stable thermal and
environmental operating conditions. The overall design builds directly
on operational experience from the CHIME instrument.

The CHORD F-engine systems, including the CRS FPGA boards together with second-stage analog amplification and filtering electronics, are distributed between two receiver huts located within the dish the array.
Each receiver hut consists of a modified 20-foot shipping container containing an internally constructed RF-shielded room providing greater than 100\,dB of shielding effectiveness. 
Thermal management for the receiver huts is provided using closed-loop chilled-water cooling systems. Cooling is supplied by external industrial water chillers similar to those commonly used in food-processing and commercial refrigeration applications. 

The GPU X-engine correlator cluster is housed separately within a
double-wide 40-foot shipping container containing an integrated RF-shielded equipment room. As with the F-engine systems, the X-engine enclosure uses chilled-water cooling to support the high thermal loads associated with GPU-based X-engine.

\section{Outrigger Design}
\label{sec:outriggers}



CHORD will be complemented by two 64-dish outrigger stations located at the Hat Creek Radio Observatory in California and the Green Bank Observatory in West Virginia. Together with the DRAO core array, these stations will provide continent-scale baselines that enable precise localization of Fast Radio Bursts and other radio sources.


Unlike the CHORD core array, the outrigger stations are not intended for 21~cm cosmology observations and do not participate in realtime FRB discovery. Their primary role is to provide high-precision localization of events detected by the core array. As a result, the outrigger design is substantially simplified relative to the core telescope. Rather than performing continuous $N^2$ correlation and transient searches, the outriggers maintain rolling baseband buffers and record data only in response to triggers generated by the DRAO core. $N^2$ correlations are recorded during bright point source transits to internally calibrate the outrigger arrays.

This simplified operational philosophy significantly reduces both computational requirements and system complexity. The current design consists of 64 dishes per station together with a reduced backend optimized for buffering, timing, and triggered recording. Since the stringent beam and systematic-control requirements of the CHORD cosmology program do not apply to the outriggers, the project is investigating the use of commercially available steel reflector antennas in place of the custom composite dishes used by the core array, following a strategy similar to that adopted by the TONE array \cite{TONEArray}. This approach offers the potential to construct the pathfinder arrays in parallel with core array  and for substantial cost savings while maintaining the performance required for transient localization.

We plan for VLBI data recordings to be triggered and correlated offline~\cite{Leung2021SynopticVLBI} with the core-array recordings to determine source positions with very high angular accuracy. 
CHIME, which operates over a narrower fractional bandwidth with lower sensitivity, routinely achieves absolute positional accuracy of 50 milliarcseconds~\cite{Leung2024VLBICorrelator} using widefield calibration techniques~\cite{Andrew2024CalibratorGrid, CHIMEFRB2025FRB20250316A}. Nevertheless, CHIME still under-resolves its diffraction limit (31 mas at 600 MHz, the center of the CHIME band) due to low sensitivity and large angular separations on sky between bright VLBI calibrators. Scaling 50 milliarcseconds to CHORD's central frequency suggests 33 mas as a conservative baseline for CHORD's typical angular resolution, but the greatly increased sensitivity of CHORD and digital system upgrades relative to CHIME can overcome both of these limitations, meaning that CHORD is likely to be limited at the 1-10 milliarcsecond level. This angular resolution will make CHORD the premier facility for studying the progenitors of fast radio bursts and similar radio transients.

\section{Project Status}
\label{sec:status}



\begin{figure*}[t]
\centering
\includegraphics[width=\textwidth]{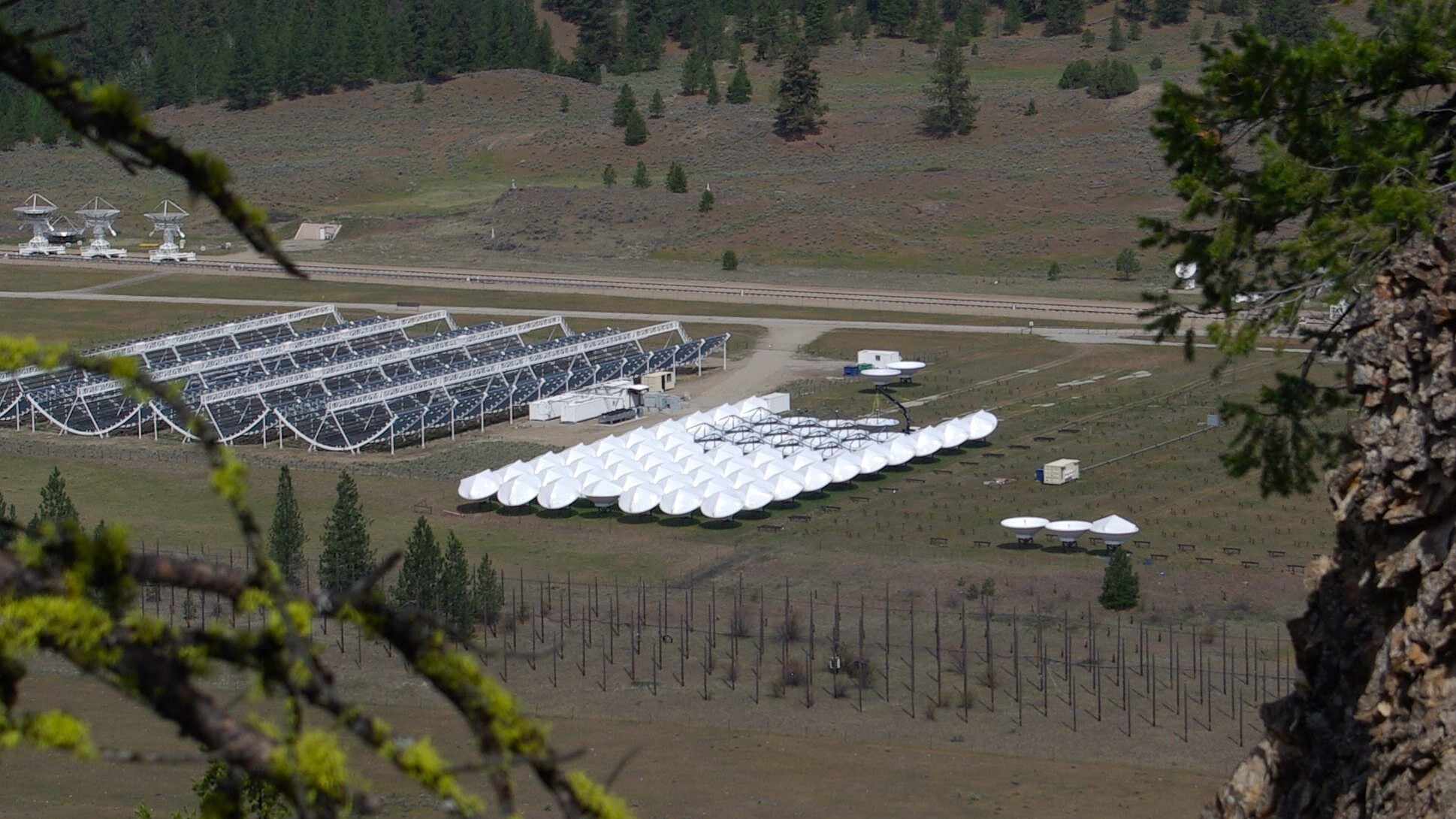}
\caption{
Photograph of the CHORD core array at the Dominion Radio Astrophysical Observatory on April 28, 2026, following installation of the 64th dish. 
Dishes with and without radomes are shown.
At the time of writing, all 64 dishes of the initial pathfinder array had been installed on their completed foundations, with integration of the analog, digital, and networking systems underway in preparation for first-light commissioning observations. The remaining dishes of the 512-element core array are planned for staged deployment through 2027, with commissioning of the full instrument expected in 2028.
Image by Brian Hoff
}
\label{fig:chord_64dish_status}
\end{figure*}

Construction activities for the CHORD core array began in September 2024 
when the first production foundation screwpiles were installed. The first dish was placed on its foundation January~16, 2025.
The production pace for dishes ramped up over the next year as staff were hired and production prodocols were optimized.
As of April~28,~2026, 64 dishes had been constructed and installed on
their completed foundations, shown in Figure~\ref{fig:chord_64dish_status}
and described in Ref.~\cite{Hoff2026Pathfinder}.
Initial commissioning activities have been carried out using a three-dish engineering sub-array used to validate the performance at the dish-level. 
A 64-dish ``pathfinder'' implementation of CHORD is scheduled for commissioning during the second and third quarters of 2026 and will serve as the first large-scale integration and science verification stage of the instrument.
Completion of the full 512-element core array is currently planned by the end of 2027, with commissioning of the complete telescope expected to follow throughout 2028.

\section{Performance from early commissioning of 3 dish array}
\label{sec:performancethreedish}



\label{sec:performance}

Initial commissioning observations have been conducted using the first three instrumented CHORD dishes, full CHORD analog chain, and the CRS F-engine backend. 
The three dishes were located along the southwestern edge of the array and formed east--west baselines of length 6.3\,m, 31.5\,m, and 37.8\,m. For these initial observations with three elements, the GPU-based X-engine was not necessary and we used the FPGA-based Pocket Correlator firmware~\cite{Hendricksen2026PocketCorrelator}. 
These observations were used to verify end-to-end system functionality, characterize the primary beam, and obtain preliminary estimates of the system temperature.

\subsection{First Fringes}

\begin{figure}[t]
\centering
\includegraphics[width=0.8\textwidth]{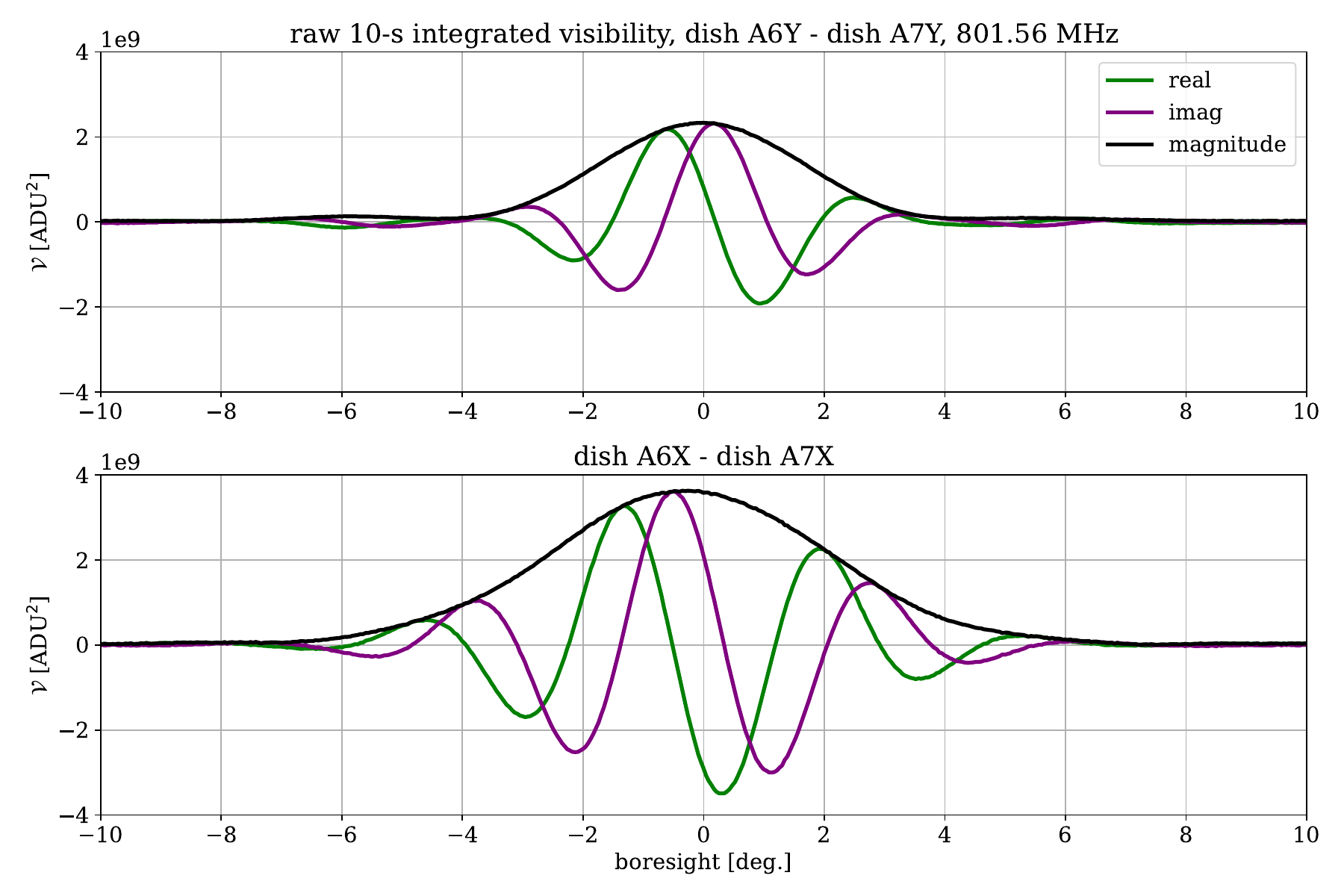}
\caption{
Interferometric fringes from early commissioning data measured with the initial three-dish CHORD array during an observation of Cassiopeia~A on October 29, 2025, demonstrating successful end-to-end integration of the CHORD signal chain.
The top panel shows the raw visibilities for the y-polarization (aligned north-south) of adjacent dishes, whereas the bottom panel
shows the x-polarization (aligned east-west).
}
\label{fig:first_fringes}
\end{figure}

An observation of the bright supernova remnant Cassiopeia A was conducted on October 29, 2025. 
Figure \ref{fig:first_fringes} presents raw visibilities, representing the ``first fringes'' for CHORD.
The upper panel shows the raw visibilities for the $y$-polarization (aligned north-south) of adjacent dishes as well as the magnitude of the visibility that corresponds to the CHORD primary beam, whereas the bottom panel shows the $x$-polarization (aligned east-west).
Strong fringes are observed demonstrating the successful operation of the complete signal chain from feed through correlation. 

\subsection{Primary Beam Measurements}

The primary beam was characterized using observations of bright transiting calibration sources across the CHORD observing band. 
Estimates 
are obtained by first applying a Gaussian fit as a function of time (or equivalently, boresight angle) to the magnitude of the raw visibilities at each frequency separately for both polarizations to recover the beam center, amplitude, and full width at half-maximum (FWHM). Assuming that the beams of each polarization are equivalent, the visibilities of an east-west transiting source measured by each polarization represent orthogonal cuts of the beam; with the convention that the $x$ ($y$) polarization is aligned east-west (north-south), the $x$ ($y$) polarization is aligned along the E-plane (H-plane). The raw visibilities are normalized by dividing the visibilities by their corresponding best-fit Gaussian amplitudes. An Airy model is then fit to the normalized visibilities to recover an additional estimate of the FWHM of the primary beam to compare with the Gaussian fits. Finally, simulations of the CHORD feed and dish were generated with the 3D EM analysis software package 
CST Studio Suite
in $5$-MHz steps across the CHORD $300$--$1500$~MHz science band in order to compare with beam shape estimates recovered from on-sky data. The FWHM of the simulations are then extracted by applying a linear interpolation to the simulated data to determine the FWHM points as a function of frequency.

\begin{figure}[t]
\centering
\includegraphics[width=0.9\textwidth]{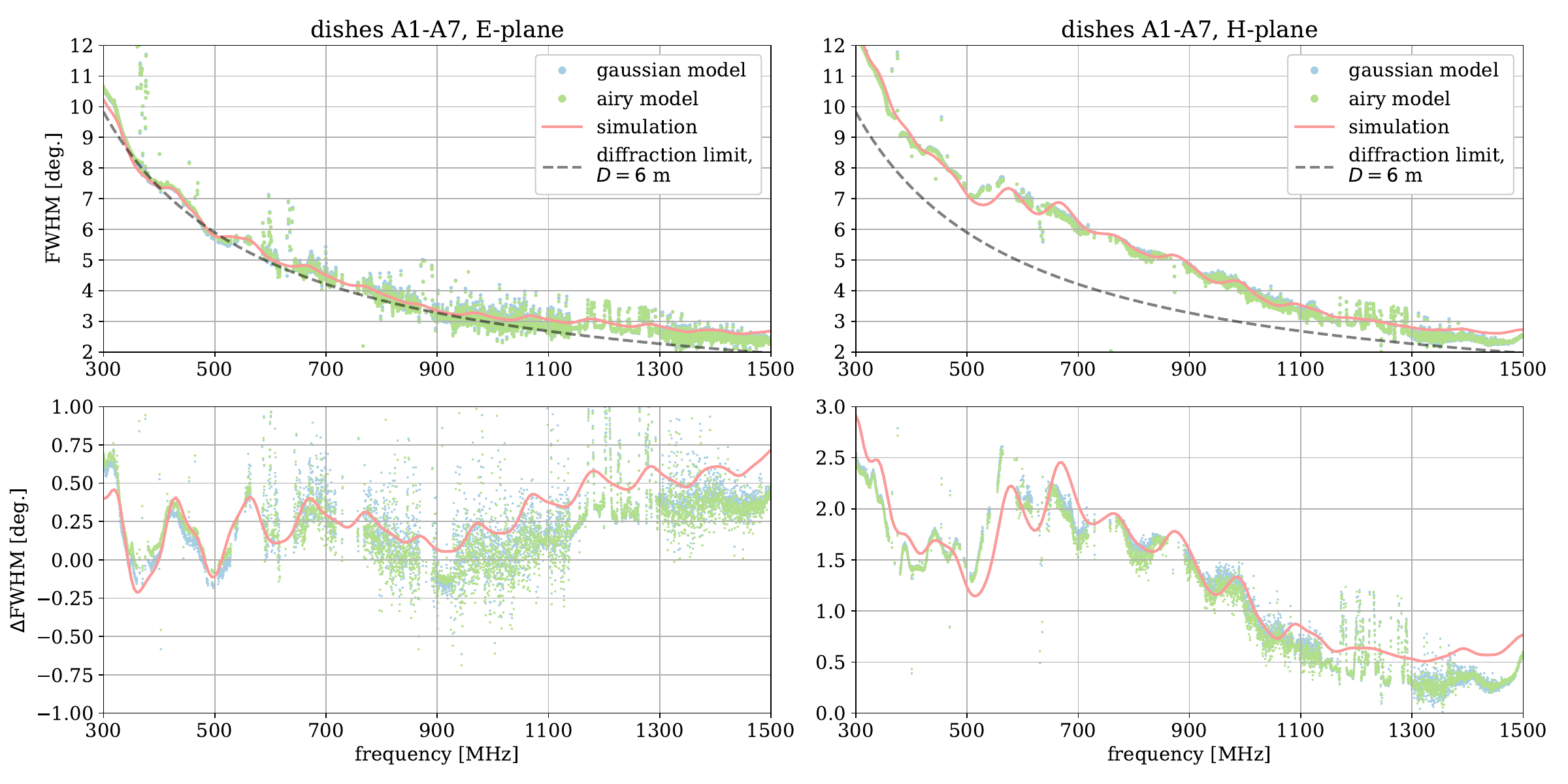}
\caption{
Measured primary-beam full width at half maximum (FWHM) as a function of frequency for the 37.8\,m baseline. Results are shown for both  polarizations together with electromagnetic simulations and the diffraction limit for a 6\,m reflector. 
The measured beam widths closely follow the simulated performance confirming that the optical performance is consistent with design expectations.
}
\label{fig:fwhm}
\end{figure}

The upper panels of Figure~\ref{fig:fwhm} present the best-fit FWHM as a function of frequency for the E- and H-planes of the longest baseline, with the Gaussian model shown in blue and the Airy model shown in green. The FWHM as estimated from CST simulations is shown in red, and the FWHM of a diffraction-limited parabolic dish with a diameter of $D = 6$~m is shown as a dashed black line for comparison. The lower panels show the difference between the FWHM estimates and the diffraction limit for easier comparison between the different estimates. The rippling pattern is due to standing waves in free space between the dish and the feed. The E-plane cut essentially achieves the diffraction limit for most frequencies, only departing slightly by $\sim 0.5$~degrees towards higher frequencies. As expected, the H-plane is $\sim 1$--$2.5$ degrees larger than the diffraction limit at frequencies below $1100$~MHz, but reaches about $0.5$~degrees above the diffraction limit above this frequency. The overall behavior of each cut is in agreement with expectations: given that the Vivaldi feeds are an extension of a simple dipole, the E-plane is narrower than the H-plane. The Gaussian and Airy models are generally consistent, but there is a notable difference at frequencies below $550$~MHz, where the FWHM estimates from the Gaussian model are lower than that of the Airy model. The first sidelobes are visibly diminished for the E-plane cuts at these frequencies, suggesting that a Gaussian is a better description of the E-plane below $550$~MHz. The overall behavior of the FWHM estimates from on-sky observations are in excellent agreement with those of the simulations, with small ($\lesssim 0.5$~degree) differences visible for certain features.

\subsection{System Temperature}

Preliminary estimates of the system temperature were obtained using calibrated visibility measurements together with simulated beam solid angles derived from electromagnetic models of the feed and reflector.
The noise in the source flux is estimated from the radiometer equation by first selecting a region of the sky far from the calibrator source, where the noise is calculated using the von Neumann Mean-Squared Successive Difference Estimator (MSSD) \cite{MSSD}. The MSSD is a robust estimator against slowly-drifting means (i.e., fringing) in the calibrated visibilities, which can be present even far from the source due to side lobes, especially at lower frequencies.
To determine the beam solid angle, we numerically integrate the simulated beam using \texttt{UVBeam} over $4 \pi$. We do not use estimates of the beam solid angle obtained from the measured visibilities, since they are not sensitive enough to recover information about the side lobes, which make important contributions to the total beam solid angle, and thus $T_\text{sys}$.

\begin{figure}[t]
\centering
\includegraphics[width=0.9\textwidth]{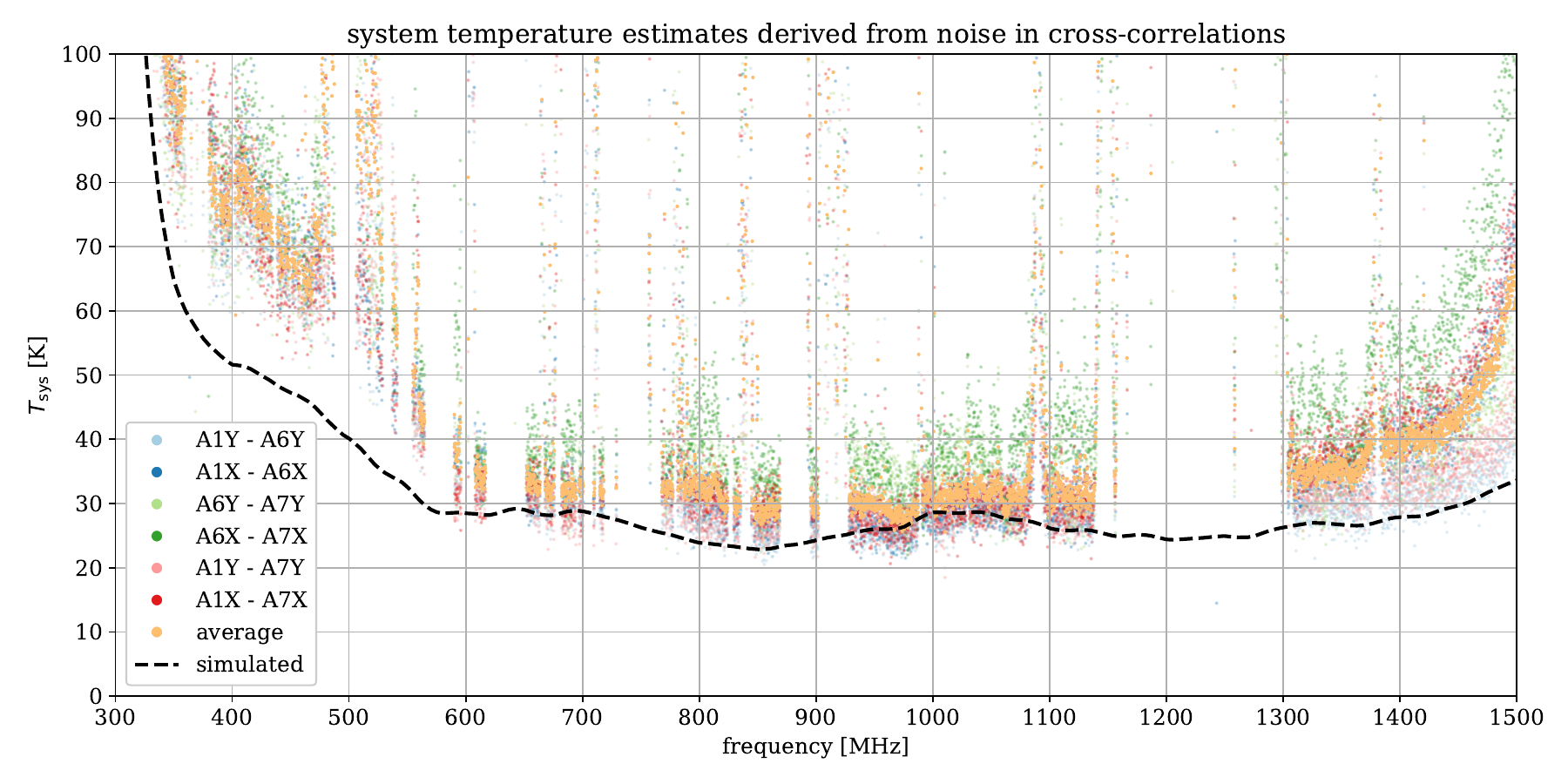}
\caption{
Preliminary estimates of the CHORD system temperature derived from
cross-correlation measurements using the three-dish engineering array.
Individual baseline and polarization measurements are shown together
with their average and the predicted performance from receiver
simulations. The measured system temperature is broadly consistent with
expectations across much of the observing band, with particularly good
agreement between 600 and 1100\,MHz.
}
\label{fig:Tsys}
\end{figure}

Figure~\ref{fig:Tsys} shows estimates of $T_\text{sys}$ for each of the measured visibilities, as well as their average in orange. 
For comparison, a simulated $T_\text{sys}$ is shown as a dashed black line, reproduced from Figure~$12$ of MacKay \emph{et al.}\cite{MacKay2023Feed}.
 The average system temperature is generally consistent with simulations, especially from $600$--$1100$~MHz, where the difference with simulations is of order $\lesssim 5$~K. Departures from expectations are evident outside this region, which are the subject of ongoing analysis. The two polarizations for the dish A6-A7 baseline are notably higher, which is most likely due to additional contributions to the measured visibilities from the Galaxy due to its shorter baseline; the $T_\text{sys}$ estimates of the longer baselines are more consistent with the simulated $T_\text{sys}$. These preliminary results demonstrate the initial success for the CHORD Pathfinder, with the performance of the first instrumented dishes largely within expectations.

\section{Conclusions}
\label{sec:conclusions}

CHORD is a next-generation wideband radio interferometer designed to address key science goals in 21\,cm cosmology, fast radio transients, pulsar astronomy,  spectral line galaxies,
and very-long-baseline interferometry. The instrument combines a highly redundant 512-dish core array with ultra-wideband feeds, direct RF sampling using RFSoC-based digital backends, and flexible realtime processing pipelines optimized for both precision mapping and time-domain astronomy.

Construction of the first 64-dishes for the pathfinder have been completed at the DRAO site, and initial commissioning observations using the first three instrumented dishes have successfully demonstrated beam performance consistent with electromagnetic simulations and system temperatures broadly in line with design expectations. These early results validate key aspects of the CHORD telescope architecture and provide confidence in the scalability of the design to the full instrument.

Installation of feeds, alignment, commissioning and early science operations of the 64-dish pathfinder will continue while construction of the complete 512-dish core array and two 64-dish outrigger stations proceeds toward completion. Upon full deployment, CHORD will serve as a powerful instrument for precision cosmology and exploration of the dynamic radio sky.
\section{Acknowledgments}
\label{sec:acknowledgments}

The CHORD core array is situated on the ancestral and unceded territory of the Syilx (Okanagan) Nation.
The CHORD collaboration recognizes the enduring stewardship of these lands by the Syilx people and are grateful for the opportunity to carry out scientific research in this region.

The CHORD collaboration acknowledges the Dominion Radio
Astrophysical Observatory (DRAO), operated by the National
Research Council Canada (NRC), as a full institutional partner
in the CHORD project and host of the CHORD core
array on its radio-protected site in Penticton, British Columbia.

CHORD infrastructure is supported through grants from the Canada Foundation for Innovation (CFI), the provinces of Alberta,
Ontario, and Quebec, the Istituto Nazionale di Astrofisica in Italy,
and participating partner institutions.
Offline computational and data-storage resources are provided by the
Digital Research Alliance of Canada.

Individual members of the collaboration acknowledge support from a variety of national and international funding programs, including the Natural Sciences and Engineering Research Council of Canada (NSERC), the Canada Research Chairs Program, the Canada Foundation for Innovation (CFI), Mitacs, the Trottier Space Institute, the National Science Foundation (NSF), the NASA Hubble Fellowship Program, the Government of Ontario, the Government of Canada, the Province of Quebec, the Province of Ontario, the French government through the France 2030 investment plan and Initiative d'Excellence d'Aix-Marseille Universit\'e (A*MIDEX), and the European Union NextGenerationEU program through Italy's National Recovery and Resilience Plan (PNRR).


\bibliography{report} 
\bibliographystyle{spiebib} 

\end{document}

%% file: chord_spie_authors_affiliations.tex
\author{The CHORD Collaboration}

\author[a,b]{Kevin Bandura}
\author[c]{Leonid Belostotski}
\author[d,e]{Lindsay Berkhout}
\author[f,g,h]{Gianni Bernardi}
\author[i]{Akanksha Bij}
\author[j]{Duncan Cameron-Steinke}
\author[d,e]{Arnab Chakraborty}
\author[d,e]{Hsin Cynthia Chiang}
\author[d,e]{Jean-Francois Cliche}
\author[k,l]{Sophia Da Costa}
\author[d,e]{Evan Davies-Velie}
\author[d,e]{Matt Dobbs}
\author[a,b]{Emmanuel Fonseca}
\author[m]{Simon Foreman}
\author[i]{Qwin Goodwin}
\author[d,e]{Ian Hendricksen}
\author[d,e,n,o]{Jason Hessels}
\author[j,p]{Alex S Hill}
\author[p]{Mohammad Islam}
\author[d,e]{Michael Jafs}
\author[d,e]{Aditya Krishna Karigiri Madhusudhan}
\author[p]{Gordon Lacy}
\author[q,r]{Dustin Lang}
\author[s]{Calvin Leung}
\author[k,t]{Jessie Lin}
\author[d,e]{Adrian Liu}
\author[u]{Vincent MacKay}
\author[u,v]{Kiyoshi W. Masui}
\author[k,l]{Juan Mena-Parra}
\author[w,x]{James Mertens}
\author[y]{Daniele Michilli}
\author[z]{Kenzie Nimmo\thanks{NASA Hubble Fellow}}
\author[w,q]{Robert Pascua}
\author[aa]{Maura Pilia}
\author[aa]{Andrea Possenti}
\author[y]{Aniket Prasad}
\author[k]{Andre Renard}
\author[d,e]{Sophia D'Agostino Rubens}
\author[y]{Mawson Sammons}
\author[q,r]{Erik Schnetter}
\author[ab]{Paul Scholz}
\author[ac,q,d,e]{Seth Siegel}
\author[d]{Jonathan Sievers}
\author[q]{Kendrick Smith}
\author[i]{Kristine Spekkens}
\author[d,e]{Shronim Tiwari}
\author[ad]{Martin Topinka}
\author[d,e,aa]{Matteo Trudu}
\author[k,l]{Keith Vanderlinde}
\author[d,e]{Dallas Wulf}
\author[k]{Yifan Zhao}

\affil[a]{Department of Physics and Astronomy, West Virginia University, Morgantown, WV 26506, USA}
\affil[b]{Center for Gravitational Waves and Cosmology, West Virginia University, Morgantown, WV 26506, USA}
\affil[c]{Department of Electrical and Software Engineering, University of Calgary, 2500 University Drive NW, Calgary, AB T2N 1N4, Canada}
\affil[d]{Department of Physics, McGill University, 3600 rue University, Montreal, QC H3A 2T8, Canada}
\affil[e]{Trottier Space Institute at McGill, McGill University, 3550 University Street, Montreal, QC H3A 2A7, Canada}
\affil[f]{INAF -- Istituto di Radioastronomia, Via Piero Gobetti 101, 40129 Bologna, Italy}
\affil[g]{Department of Physics and Electronics, Rhodes University, Makhanda 6140, South Africa}
\affil[h]{South African Radio Astronomy Observatory, 2 Fir Street, Black River Park, Observatory, Cape Town 7925, South Africa}
\affil[i]{Department of Physics, Engineering Physics and Astronomy, Queen's University, Kingston, ON K7L 3N6, Canada}
\affil[j]{Department of Physics and Astronomy, University of British Columbia, Okanagan Campus, Kelowna, BC V1V 1V7, Canada}
\affil[k]{Dunlap Institute for Astronomy \& Astrophysics, University of Toronto, 50 St. George Street, Toronto, ON M5S 3H4, Canada}
\affil[l]{David A. Dunlap Department of Astronomy \& Astrophysics, University of Toronto, 50 St. George Street, Toronto, ON M5S 3H4, Canada}
\affil[m]{School of Earth and Space Exploration, Arizona State University, Tempe, AZ 85287, USA}
\affil[n]{Anton Pannekoek Institute for Astronomy, University of Amsterdam, 1098 XH Amsterdam, The Netherlands}
\affil[o]{ASTRON, Netherlands Institute for Radio Astronomy, 7991 PD Dwingeloo, The Netherlands}
\affil[p]{Dominion Radio Astrophysical Observatory, Herzberg Astronomy and Astrophysics Research Centre, National Research Council Canada, 717 White Lake Road, Penticton, BC V2A 6J9, Canada}
\affil[q]{Perimeter Institute for Theoretical Physics, 31 Caroline Street North, Waterloo, ON N2L 2Y5, Canada}
\affil[r]{Department of Physics and Astronomy, University of Waterloo, Waterloo, ON N2L 3G1, Canada}
\affil[s]{Department of Astronomy, University of California, Berkeley, Berkeley, CA 94720, USA}
\affil[t]{Department of Physics, University of Toronto, 60 St. George Street, Toronto, ON M5S 1A7, Canada}
\affil[u]{MIT Kavli Institute for Astrophysics and Space Research, Massachusetts Institute of Technology, 77 Massachusetts Avenue, Cambridge, MA 02139, USA}
\affil[v]{Department of Physics, Massachusetts Institute of Technology, 77 Massachusetts Avenue, Cambridge, MA 02139, USA}
\affil[w]{University of Toronto, Toronto, ON M5S 1A1, Canada}
\affil[x]{Department of Astronomy, Case Western Reserve University, Cleveland, OH 44106, USA}
\affil[y]{Laboratoire d'Astrophysique de Marseille, Aix Marseille Univ, CNRS, CNES, LAM, Marseille, France}
\affil[z]{Center for Interdisciplinary Exploration and Research in Astrophysics, Northwestern University, Evanston, IL 60208, USA}
\affil[aa]{INAF -- Osservatorio Astronomico di Cagliari, Via della Scienza 5, 09047 Selargius, Italy}
\affil[ab]{York Centre for Astrophysics, Department of Physics and Astronomy, York University, 4700 Keele Street, Toronto, ON M3J 1P3, Canada}
\affil[ac]{SKA Observatory, Jodrell Bank, Lower Withington, Macclesfield SK11 9FT, United Kingdom}
\affil[ad]{Institute of Theoretical Physics, Faculty of Mathematics and Physics, Charles University, Prague, Czech Republic}

%% file: report.bib
@article{CHIMEOverview2022,
  author = {{CHIME Collaboration}},
  title = {An Overview of {CHIME}, the {C}anadian {H}ydrogen {I}ntensity {M}apping {E}xperiment},
  journal = {Astrophysical Journal},
  volume = {863},
  pages = {48},
  year = {2022},
  doi = {10.3847/1538-4357/ac6fd9}
}

@article{CHORDWhitepaper2019,
  author = {Vanderlinde, K. and Bandura, K. and Belostotski, L. and others},
  title = {The {Canadian} {Hydrogen} {Observatory} and {Radio-transient} {Detector} {(CHORD)}},
  journal = {Long Range Plan 2020 White Paper},
  year = {2019},
  volume = {arXiv:1911.01777},
  eprint = {arXiv:1911.01777}
}

@article{CHIMEFRB2019,
  author = {{CHIME/FRB Collaboration}},
  title = {Observations of {F}ast {R}adio {B}ursts at Frequencies down to 400 {M}egahertz},
  journal = {Nature},
  volume = {566},
  pages = {230--234},
  year = {2019},
  doi = {10.1038/s41586-018-0867-7}
}

@ARTICLE{CHIMECatalog1,
       author = {{CHIME/FRB Collaboration}},
        title = "{The First CHIME/FRB Fast Radio Burst Catalog}",
      journal = {Astrophysical Journal Supplement Series},
         year = 2021,
        month = dec,
       volume = {257},
       number = {2},
          eid = {59},
        pages = {59},
          doi = {10.3847/1538-4365/ac33ab},
archivePrefix = {arXiv},
       eprint = {2106.04352},
 primaryClass = {astro-ph.HE},
       adsurl = {https://ui.adsabs.harvard.edu/abs/2021ApJS..257...59C}
}

@article{CHIMECatalog2,
  author = {{CHIME/FRB Collaboration}},
  title = {The Second {CHIME/FRB} {Fast} {Radio} {Burst} Catalog},
  journal = {arXiv e-prints},
  year = {2026},
  volume = {arXiv:2601.09399},
  eprint = {2601.09399},
  archivePrefix = {arXiv},
  primaryClass = {astro-ph.HE}
}

@article{CHIMEAuto21cm2025,
  author = {{CHIME Collaboration}},
  title = {First Detection of the Cosmological 21 cm Intensity Mapping Signal in Auto-correlation at Redshift 1},
  journal = {arXiv e-prints},
  year = {2025},
  volume = {2511.19620},
  eprint = {2511.19620},
  archivePrefix = {arXiv},
  primaryClass = {astro-ph.CO}
}

@article{Bandura2016,
  author = {Bandura, K. and others},
  title = {{Canadian} {Hydrogen} {Intensity} {Mapping} {Experiment} Digital Backend},
  journal = {Journal of Astronomical Instrumentation},
  volume = {5},
  number = {4},
  pages = {1641005},
  year = {2016}
}

@misc{chFPGA,
  author = {Cliche, J-F. and others},
  title = {{chFPGA} Firmware Documentation},
  howpublished = {\url{http://winterlandcosmology.bitbucket.io/chfpga/}},
  year = {2025}
}

@mastersthesis{HendricksonThesis,
  title        = {{Observational, digital readout, and calibration techniques for studying the redshifted 21cm signal of hydrogen}},
  author       = {Ian Hendricksen},
  year         = 2023,
  address      = {Montréal, QC, Canada},
  note         = {Available at \url{https://escholarship.mcgill.ca/concern/theses/2b88qj88r?locale=en}},
  school       = {McGill University},
  type         = {{Master's thesis}}
}

@inproceedings{Montgomery2024CRS,
  author = {Montgomery, Joshua and Avelino, Wellington and Dobbs, Matt and Letang, Joseph and Rouble, Maclean and Savchyn, Sofiia and Smecher, Graeme},
  title = {The {CRS}: a scalable full-stack control system for Microwave Kinetic Inductance Detectors},
  booktitle = {Millimeter, Submillimeter, and Far-Infrared Detectors and Instrumentation for Astronomy XII},
  series = {Proceedings of SPIE},
  volume = {13102},
  pages = {131020R},
  year = {2024},
  doi = {10.1117/12.3019425}
}

@misc{CASPER,
  author = {{CASPER Collaboration}},
  title = {CASPER: Collaboration for Astronomy Signal Processing and Electronics Research},
  howpublished = {\url{https://casper.berkeley.edu/}},
  year = {2025}
}

@article{MacKay2023Feed,
  author = {MacKay, Vincent and Lai, Mark and Shmerko, Peter and Wulf, Dallas and Belostotski, Leonid and Vanderlinde, Keith},
  title = {Low-Cost, Low-Loss, Ultra-Wideband Compact Feed for Interferometric Radio Telescopes},
  journal = {Journal of Astronomical Instrumentation},
  year = {2023},
         volume = {12},
       number = {4},
  doi = {10.1142/S2251171723500083},
  eprint = {2210.07477},
  archivePrefix = {arXiv},
  primaryClass = {astro-ph.IM}
}

@misc{Bij2026,
  author = {Bij, Akanksha and {The CHORD Collaboration}},
  title = {{CHORD} {HI}-Galaxy Survey Forecasts: {Searching} for nearby dark galaxies and high redshift giants},
  journal = {in prep},
  year = {2026}
}

@inproceedings{Islam2022D3A6,
  author = {Islam, Mohammad Nouroz and {\"O}l{\c{c}}ek, Deniz and Lacy, Gordon and Chiang, Hsin Cynthia and Hellyer, Richard},
  title = {Errors in {Deep} {Dish} {Development} {Array} (6m) Construction and Metrology Steps},
  booktitle = {Ground-based and Airborne Telescopes IX},
  series = {Proceedings of SPIE},
  volume = {12182},
  pages = {121820F},
  year = {2022},
  doi = {10.1117/12.2629423}
}

@inproceedings{Islam2020D3A,
  author = {Islam, Mohammad Nouroz and Hellyer, Richard and Chiang, Hsin Cynthia and Vanderlinde, Keith},
  title = {Development of Composite Reflector Technology for the {Deep} {Dish} {Development} {Array}},
  booktitle = {Ground-based and Airborne Telescopes VIII},
  series = {Proceedings of SPIE},
  volume = {11445},
  pages = {1144523},
  year = {2020},
  doi = {10.1117/12.2562068}
}

@inproceedings{Hoff2026Pathfinder,
  author = {Hoff, Brian and Wulf, Dallas and Islam, Mohammad and Dueck, Tayron and Hellyer, Richard},
  title = {Dish Working Group Progress Toward the CHORD 512-Dish Core Array: Completion of the 64-Dish Pathfinder},
  booktitle = {Radio Telescopes, Technologies, and Methods},
  series = {Proceedings of SPIE},
  volume = {14153},
  pages = {14153-62},
  year = {2026}
}

@inproceedings{DallasSPIE2026ProjManagement,
  author = {Wulf, Dallas and Hoff, Brian and Dobbs, Matt and Vanderline, Keith},
  title = {Project management of the Canadian Hydrogen Observatory and Radio-transient Detector},
  booktitle = {Radio Telescopes, Technologies, and Methods},
  series = {Proceedings of SPIE},
  volume = {14152},
  pages = {14152-4},
  year = {2026}
}

@article{MSSD,
 ISSN = {00034851},
 URL = {http://www.jstor.org/stable/2235765},
 author = {J. von Neumann and R. H. Kent and H. R. Bellinson and B. I. Hart},
 journal = {The Annals of Mathematical Statistics},
 number = {2},
 pages = {153--162},
 publisher = {Institute of Mathematical Statistics},
 title = {The Mean Square Successive Difference},
 urldate = {2026-05-29},
 volume = {12},
 year = {1941}
}

@article{Shaw:2014khi,
    author = "Shaw, J. Richard and Sigurdson, Kris and Sitwell, Michael and Stebbins, Albert and Pen, Ue-Li",
    title = "{Coaxing Cosmic 21 cm Fluctuations from the Polarized Sky Using $m$-Mode Analysis}",
    eprint = "1401.2095",
    archivePrefix = "arXiv",
    primaryClass = "astro-ph.CO",
    reportNumber = "FERMILAB-PUB-14-044-A",
    doi = "10.1103/PhysRevD.91.083514",
    journal = "Phys. Rev. D",
    volume = "91",
    number = "8",
    pages = "083514",
    year = "2015"
}

@article{CHIME:2025cee,
  author = {{CHIME Collaboration}},
  title = {Detection of the Cosmological 21 cm Signal in Auto-correlation at $z \sim 1$ with the Canadian Hydrogen Intensity Mapping Experiment},
  journal = {arXiv e-prints},
  year = {2025},
    volume = {arXiv:2511.19620},
    eprint = "2511.19620",
    archivePrefix = "arXiv",
    primaryClass = "astro-ph.CO"
}

@misc{hans,
      title={Matched Filtering for the {Canadian} {Hydrogen} {Observatory} and {Radio-transient} {Detector} Galaxy Search}, 
      author={Hans S. Hopkins and Dustin Lang and Kendrick Smith and Kristine Spekkens and Simon Foreman and Akanksha Bij},
      year={2026},
      eprint={2603.03520},
      archivePrefix={arXiv},
      primaryClass={astro-ph.IM},
      url={https://arxiv.org/abs/2603.03520} 
}

@article{Rafiei-Ravandi:2022rwl,
    author = "Rafiei-Ravandi, Masoud and Smith, Kendrick M.",
    title = "{Mitigating Radio Frequency Interference in CHIME/FRB Real-time Intensity Data}",
    eprint = "2206.07292",
    archivePrefix = "arXiv",
    primaryClass = "astro-ph.IM",
    doi = "10.3847/1538-4365/acc252",
    journal = "Astrophys. J. Suppl.",
    volume = "265",
    number = "2",
    pages = "62",
    year = "2023"
}

@article{Taylor_2019,
   title={Spectral Kurtosis-Based {RFI} Mitigation for {CHIME}},
   volume={08},
   ISSN={2251-1725},
   url={http://dx.doi.org/10.1142/S225117171940004X},
   DOI={10.1142/s225117171940004x},
   number={01},
   journal={Journal of Astronomical Instrumentation},
   publisher={World Scientific Puab Co Pte Ltd},
   author={Taylor, Jacob and Denman, Nolan and Bandura, Kevin and Berger, Philippe and Masui, Kiyoshi and Renard, Andre and Tretyakov, Ian and Vanderlinde, Keith},
   year={2019},
   month=Mar
}

@article{Amiri_2025,
doi = {10.3847/1538-4357/adfdcc},
url = {https://doi.org/10.3847/1538-4357/adfdcc},
year = {2025},
month = {oct},
publisher = {The American Astronomical Society},
volume = {993},
number = {1},
pages = {55},
author = {{CHIME/FRB Collaboration} and Amiri, Mandana and Andersen, Bridget C. and Andrew, Shion and Bandura, Kevin and Bhardwaj, Mohit and Bhopi, Kalyani and Bidula, Vadym and Boyle, P. J. and Brar, Charanjot and Carlson, Mark and Cassanelli, Tomas and Cassity, Alyssa and Chatterjee, Shami and Cliche, Jean-FranÃ§ois and Curtin, Alice P. and Darlinger, Rachel and DeBoer, David R. and Dobbs, Matt and Dong, Fengqiu Adam and Eadie, Gwendolyn and Fonseca, Emmanuel and Gaensler, B. M. and Gusinskaia, Nina and Halpern, Mark and Hendricksen, Ian and Hessels, Jason and Joseph, Ronniy C. and Kaczmarek, Jane and Kaspi, Victoria M. and Khairy, Kholoud and Landecker, T. L. and Lanman, Adam E. and Lau, Albert Wai Kit and Lazda, Mattias and Leung, Calvin and Main, Robert A. and Masui, Kiyoshi W. and Mckinven, Ryan and Mena-Parra, Juan and Meyers, Bradley W. and Michilli, Daniele and Milutinovic, Nikola and Nimmo, Kenzie and Noble, Gavin and Pandhi, Ayush and Pearlman, Aaron B. and Peterson, Jeffrey B. and Petroff, Emily and Pleunis, Ziggy and Pollak, Alexander W. and Rafiei-Ravandi, Masoud and Renard, Andre and Sammons, Mawson W. and Sand, Ketan R. and Sanghavi, Pranav and Scholz, Paul and Shah, Vishwangi and Shin, Kaitlyn and Siegel, Seth R. and Siemion, Andrew and Sievers, Jonathan L. and Smith, Kendrick and Spear, David and Stairs, Ingrid and Vanderlinde, Keith and Wang, Haochen and Willis, Jacob P. and Zegmott, Tarik J.},
title = {{CHIME/FRB} {Outriggers:} {Design} Overview},
journal = {The Astrophysical Journal},
}

@article{Leung2024VLBICorrelator,
  author = {Leung, Calvin and Andrew, Shion and Masui, Kiyoshi W. and Brar, Charanjot and Cassanelli, Tomas and Chatterjee, Shami and Kaspi, Victoria and Khairy, Kholoud and Lanman, Adam E. and Lazda, Mattias and Mena-Parra, Juan and Noble, Gavin and Pearlman, Aaron B. and Rahman, Mubdi and Sanghavi, Pranav and Shah, Vishwangi},
  title = {A {VLBI} Software Correlator for Fast Radio Transients},
  journal = {arXiv e-prints},
  year = {2024},
  volume = {arXiv:2403.05631},
  eprint = {2403.05631},
  archivePrefix = {arXiv},
  primaryClass = {astro-ph.IM},
  doi = {10.48550/arXiv.2403.05631}
}

@article{CHIMEFRB2025FRB20250316A,
  author = {{CHIME/FRB Collaboration}},
  title = {FRB 20250316A: A Brilliant and Nearby One-Off {Fast} {Radio} {Burst} Localized to 13 Parsec Precision},
  journal = {arXiv e-prints},
  year = {2025},
  volume = {arXiv:2506.19006},
  eprint = {2506.19006},
  archivePrefix = {arXiv},
  primaryClass = {astro-ph.HE},
  doi = {10.48550/arXiv.2506.19006}
}

@article{Leung2021SynopticVLBI,
  author = {Leung, Calvin and Mena-Parra, Juan and Masui, Kiyoshi and Bhardwaj, Mohit and Boyle, P. J. and Brar, Charanjot and Bruneault, Mathieu and Cassanelli, Tomas and Cubranic, Davor and Kaczmarek, Jane F. and Kaspi, Victoria and Landecker, Tom and Michilli, Daniele and Milutinovic, Nikola and Patel, Chitrang and Renard, Andre and Sanghavi, Pranav and Scholz, Paul and Stairs, Ingrid H. and Vanderlinde, Keith},
  title = {A Synoptic {VLBI} Technique for Localizing Non-Repeating {Fast} {Radio} {Bursts} with {CHIME/FRB}},
  journal = {Astronomical Journal},
  volume = {161},
  number = {2},
  pages = {81},
  year = {2021},
  doi = {10.3847/1538-3881/abd174},
  eprint = {2008.11738},
  archivePrefix = {arXiv},
  primaryClass = {astro-ph.IM}
}

@article{Andrew2024CalibratorGrid,
  author = {Andrew, Shion and Leung, Calvin and Li, Alexander and Masui, Kiyoshi W. and Andersen, Bridget C. and Bandura, Kevin and Curtin, Alice P. and Kaczmarek, Jane and Lanman, Adam E. and Lazda, Mattias and Mena-Parra, Juan and Michilli, Daniele and Nimmo, Kenzie and Pearlman, Aaron B. and Rahman, Mubdi and Shah, Vishwangi and Shin, Kaitlyn and Wang, Haochen},
  title = {A {VLBI} Calibrator Grid at 600 {MHz} for Fast Radio Transient Localizations with {CHIME/FRB} {Outriggers}},
  journal = {arXiv e-prints},
  year = {2024},
  volume = {arXiv:2409.11476},
  eprint = {2409.11476},
  archivePrefix = {arXiv},
  primaryClass = {astro-ph.IM},
  doi = {10.48550/arXiv.2409.11476}
}

@ARTICLE{10124757,
  author={Lai, Mark and Mackay, Vincent and Wulf, Dallas and Shmerko, Peter and Belostotski, Leonid},
  journal={IEEE Microwave and Wireless Technology Letters}, 
  title={0.3–1.5-{GHz} {LNA} With Wideband Noise and Power Matching for Radio Astronomy}, 
  year={2023},
  volume={33},
  number={8},
  pages={1163-1166},
  doi={10.1109/LMWT.2023.3272211}}

@INPROCEEDINGS{Shaw2014,
       author = {{Shaw}, J. Richard and {Sigurdson}, Kris and {Pen}, Ue-Li and {Stebbins}, Albert and {Sitwell}, Michael},
        title = "{Analysing transit telescopes with the m-mode formalism}",
    booktitle = {2014 USNC-URSI Radio Science Meeting (Joint with AP-S Symposium)},
         year = 2014,
        month = jan,
    publisher = {IEEE},
          eid = {179},
        pages = {179},
          doi = {10.1109/USNC-URSI-NRSM.2014.6928117},
       adsurl = {https://ui.adsabs.harvard.edu/abs/2014usnc.conf..179S}
}

@ARTICLE{CordesChatterjee2019,
       author = {{Cordes}, James M. and {Chatterjee}, Shami},
        title = "{Fast Radio Bursts: An Extragalactic Enigma}",
      journal = {Annual Review of Astronomy and Astrophysics},
         year = 2019,
        month = aug,
       volume = {57},
        pages = {417-465},
          doi = {10.1146/annurev-astro-091918-104501},
archivePrefix = {arXiv},
       eprint = {1906.05878},
 primaryClass = {astro-ph.HE},
       adsurl = {https://ui.adsabs.harvard.edu/abs/2019ARA&A..57..417C}
}

@article{Kotekan,
  author = {Renard, Andre and Shaw, Richard and Ng, Cherry and Nitsche, Rick and Taylor, Jacob and Vanderlinde, Keith and Masui, Kiyoshi and Willis, James and Fandino, Mateus and Cubranic, Davor and Naidu, Arun and Bandura, Kevin and Xu, Jintao and Connor, Liam and Brar, Shiny and Kefala, Anja and Fonseca, Emmanuel and Lanman, Adam and Tretyakov, Ian},
  title = {kotekan: High Performance Radio Data Processing Pipeline},
  journal = {Astrophysics Source Code Library},
  year = {2025},
  month = apr,
  volume = {ascl:2504.030}, 
  note = {\url{https://ascl.net/2504.030}}
}

@ARTICLE{TONEArray,
       author = {{Sanghavi}, Pranav and {Leung}, Calvin and {Bandura}, Kevin and {Cassanelli}, Tomas and {Kaczmarek}, Jane and {Kaspi}, Victoria M. and {Khairy}, Kholoud and {Lanman}, Adam and {Lazda}, Mattias and {Masui}, Kiyoshi W. and {Mena-Parra}, Juan and {Michilli}, Daniele and {Pen}, Ue-Li and {Peterson}, Jeffrey B. and {Rahman}, Mubdi and {Shah}, Vishwangi},
        title = "{TONE: A CHIME/FRB Outrigger Pathfinder for Localizations of Fast Radio Bursts using Very Long Baseline Interferometry}",
      journal = {Journal of Astronomical Instrumentation},
         year = 2024,
        month = jan,
       volume = {13},
       number = {3},
          eid = {2450010-589},
        pages = {2450010-589},
          doi = {10.1142/S2251171724500107},
archivePrefix = {arXiv},
       eprint = {2304.10534},
 primaryClass = {astro-ph.IM},
       adsurl = {https://ui.adsabs.harvard.edu/abs/2024JAI....1350010S}
}

@article{Liu2019SIGGMA,
  author = {Liu, Bingzhi and Anderson, Loren D. and McIntyre, Timothy and Roshi, D. Anish and Churchwell, Ed and Minchin, Robert and Terzian, Yervant},
  title = {The Survey of Ionized Gas of the Galaxy, Made with the Arecibo Telescope (SIGGMA): Inner Galaxy Data Release},
  journal = {Astrophysical Journal Supplement Series},
  volume = {240},
  number = {2},
  pages = {14},
  year = {2019},
  doi = {10.3847/1538-4365/aaf4fd}
}

@article{Anderson2021SPICY,
  author = {Anderson, Loren D. and Wenger, Trey V. and Armentrout, William P. and Balser, Dana S. and Bania, Thomas M. and Brown, Christopher and Luisi, Matteo},
  title = {SPICY: the Spectral-line Survey of Protocluster and H II Region Candidates},
  journal = {Astrophysical Journal Supplement Series},
  volume = {254},
  number = {2},
  pages = {28},
  year = {2021},
  doi = {10.3847/1538-4365/abf986}
}

@article{Haffner2003WHAM,
  author = {Haffner, L. M. and Reynolds, R. J. and Tufte, S. L. and Madsen, G. J. and Jaehnig, K. P. and Percival, J. W.},
  title = {The Wisconsin H$\alpha$ Mapper Northern Sky Survey},
  journal = {Astrophysical Journal Supplement Series},
  volume = {149},
  number = {2},
  pages = {405--422},
  year = {2003},
  doi = {10.1086/378850}
}

@inproceedings{Islam2026PerformanceModeling,
  author = {Islam, Mohammad N. et al.},
  title = {Performance Modelling for Radio Telescopes: An Integrated Framework of Metrology, Finite Element Analysis and Electromagnetic Simulation},
  booktitle = {Radio Telescopes, Technologies, and Methods},
  series = {Proceedings of SPIE},
  volume = {14147},
  pages = {14147-175},
  year = {2026},
  organization = {SPIE}
}

@inproceedings{Islam2026CompositeReflectors,
  author = {Islam, Mohammad N. et al.},
  title = {From Concept to Production: The Making of Composite Reflectors for the Canadian Hydrogen Observatory and Radio-transient Detector},
  booktitle = {Radio Telescopes, Technologies, and Methods},
  series = {Proceedings of SPIE},
  volume = {14153},
  pages = {14153-72},
  year = {2026},
  organization = {SPIE}
}

@inproceedings{Hendricksen2026PocketCorrelator,
  author = {Hendricksen, Ian and Dobbs, Matt and Cliche, Jean-Francois and
            Schnetter, Erik and Wulf, Dallas and Islam, Mohammad and
            Bandura, Kevin and the CHORD Collaboration},
  title = {Commissioning of the CHORD F-Engine and System Validation with a Fully FPGA-Based F-X Pocket Correlator},
  booktitle = {Radio Telescopes, Technologies, and Methods},
  series = {Proceedings of SPIE},
  volume = {14153},
  pages = {14153-76},
  year = {2026},
  organization = {SPIE}
}

@misc{Grafana,
  author       = {{Grafana Labs}},
  title        = {Grafana: The Open and Composable Observability and Data Visualization Platform},
  year         = {2026},
  howpublished = {\url{https://grafana.com/}},
  note         = {Open-source software, accessed July 2026}
}

@inproceedings{Turnbull2018Prometheus,
  author    = {Brazil, Brian},
  title     = {Prometheus: Up \& Running},
  booktitle = {O'Reilly Media},
  year      = {2018},
  publisher = {O'Reilly Media},
  isbn      = {978-1492034148}
}
